\documentclass[final,onefignum,onetabnum]{siamonline190516}

\newcommand\figref[1]{Figure\ \ref{#1}}
\newcommand\secref[1]{Section\ \ref{#1}}

\newcommand\bdm[3]{\mathrm{diag}\left(\begin{bmatrix}#1&#2&#3\end{bmatrix}\right)}

\usepackage{lipsum}
\usepackage{amsfonts}
\usepackage{graphicx}
\usepackage{epstopdf}
\usepackage{algorithmic}
\usepackage[hang]{footmisc}
\usepackage{xcolor}
\usepackage{amsmath,mleftright}
\usepackage{xparse}
\usepackage{caption,subcaption,booktabs,epstopdf}

\ifpdf
  \DeclareGraphicsExtensions{.eps,.pdf,.png,.jpg}
\else
  \DeclareGraphicsExtensions{.eps}
\fi

\usepackage{enumitem}
\setlist[enumerate]{leftmargin=.5in}
\setlist[itemize]{leftmargin=.5in}

\newsiamremark{remark}{Remark}
\newsiamremark{hypothesis}{Hypothesis}
\crefname{hypothesis}{Hypothesis}{Hypotheses}
\newsiamthm{claim}{Claim}

\headers{Scalable Physics-based MLE using Hierarchical matrices}{Y. Chen, M. Anitescu}

\title{Scalable Physics-based Maximum Likelihood Estimation using Hierarchical Matrices}

\author{Yian Chen\thanks{Department of Statistics, University of Chicago, Chicago, IL
  (\email{yianc@uchicago.edu}).}
\and Mihai Anitescu\thanks{Mathematics and Computer Science Division, Argonne National Laboratory, 9600 S. Cass Ave., Lemont, IL 
  (\email{anitescu@mcs.anl.gov}).}}

\usepackage{amsopn}

\ifpdf
\hypersetup{
  pdftitle={An Example Article},
  pdfauthor={D. Doe, P. T. Frank, and J. E. Smith}
}
\fi

\begin{document}

\maketitle

\begin{abstract}
    Physics-based covariance models provide a systematic way to construct covariance models that are consistent with
    the underlying physical laws in Gaussian process analysis. The unknown parameters in the covariance models can be estimated using maximum likelihood estimation, but direct construction of the covariance matrix and classical strategies of computing with it requires $n$ physical model runs,  $n^2$ storage complexity, and $n^3$ computational complexity. To address such challenges, we propose to approximate the discretized covariance function using hierarchical matrices. By utilizing randomized range sketching for individual off-diagonal blocks, the construction process of the hierarchical covariance approximation requires $O(\log{n})$ physical model applications and the maximum likelihood computations require $O(n\log^2{n})$ effort per iteration. We propose a new approach to compute exactly the trace of products of hierarchical matrices which results in the expected Fischer information matrix being computable in $O(n\log^2{n})$  as well. The construction is totally matrix-free and the derivatives of the covariance matrix can then be approximated in the same hierarchical structure by differentiating the whole process. Numerical results are provided to demonstrate the effectiveness, accuracy, and efficiency of the proposed method for parameter estimations and uncertainty quantification.
\end{abstract}

\begin{keywords}
    Gaussian process, maximum likelihood estimation, hierarchical matrices, parameter estimation, uncertainty quantification, fast algorithms.
\end{keywords}

\begin{AMS}
    62F10, 62F25, 65F55, 15A15
\end{AMS}

\section{Introduction}\label{sec:intro}
Gaussian processes (GPs) have been widely applied throughout the statistical and machine learning communities for modeling and regression problems \cite{gp-ref-1,gp-ref-2}. For overviews of their usage, one can refer to several extensive monographs \cite{gp-in-ml,intro-gp-2,intro-gp-3}. One of the most appealing features of GPs is that they provide a fully probabilistic framework for modeling variability in stochastic processes, predicting unobserved values and providing prediction uncertainties. Gaussian process regression, also referred to as kriging, is considered one of the standard methods for modeling a range of natural phenomena from geophysics to biology \cite{intro-gp-3,stein-theory-of-kriging,gp-geo-2}. In practice, we may want to model multiple processes that are highly correlated. Extending the technique to multiple target variables is known as co-kriging \cite{stein-theory-of-kriging,co-kriging} or multi-kriging \cite{gp-ref-1}. In this work, we focus on multivariate co-kriging where the variables can be extracted from spatial processes or spatio-temporal processes.

An important feature of GPs is that the model can be fully characterized by its mean and covariance functions. Correct and accurate covariance models play an essential role in setting up meaningful GP regression problems. When applying kriging to single random fields, one can construct or choose proper parametric covariance functions based on a variety of practical or theoretical guidelines \cite{gp-in-ml,stein-theory-of-kriging}. When modeling multiple processes, however, it is hard to construct correct covariance structure that accounts for how the processes interact. To tackle the difficulty, many efforts have been carried out to incorporate prior knowledge about the underlying physical principles of the processes into the covariance model structure. A typical strategy of including prior knowledge of a real system governed
by known physical laws is hierarchical Bayesian modeling \cite{bayesian-1,bayesian-2}. Examples include atmospheric modeling \cite{atom-1,atmo-2,atmo-3,atmo-4}
and environmental sciences \cite{atmo-2,env-1,env-2,env-3,env-4}. Recently in \cite{physics-based-cov-model}, the authors proposed a systematic framework of assembling consistent auto- and cross-covariance models based on known physical processes. The endeavor showed that significant improvements in the forecast efficiency can be achieved by using physics-based covariance models. To address the complexity issue for large scale problems, our previous work \cite{scalable-physics-gp} propose a way to construct the low-rank approximations of the physics-based covariance models using black-box forward solvers for implicit physical models. It results in a scalable approach of performing GP analysis using physics-based covariance models. However the low-rank assumption may not hold in many applications where the covariance model is often of full-rank. In this work, we propose to relax the assumption and approximate the physics-based covariance models in more general hierarchical matrix format while still only relying on forward applications of the physical models and maintaining a quasilinear complexity. We include more details about the physics-based covariance models in \secref{sec:cov model and MLE}.

Another feature of maximum likelihood computation for  GPs is that the main step in the calculation requires computing the inverse and determinant of the covariance matrix, in turn typically carried out by Cholesky factorization.  If the matrix has no exploitable structure, the Cholesky factorization requires $O(n^3)$ computations and $O(n^2)$ storage for an $n\times n$ matrix. With the dramatic increase in the efficiency and capability of data collection, the computational and memory cost can soon become prohibitive for large scale problems. As a result, devising approximations to reduce both the computational and memory cost became a burgeoning field, using, for example, low-rank methods \cite{low-rank-1,low-rank-2}, matrix tapering \cite{mat-tapering-1,mat-tapering-2,mat-tapering-3}, Gaussian predictive processes \cite{gaussian-predictive}, fixed-rank kriging \cite{fixed-rank-kriging}, and Gaussian Markov random field approximations \cite{GMRF-1,GMRF-2,GMRF-3}. A matrix-free approach
for solving the multi-parametric Gaussian maximum likelihood problem was developed in \cite{matrix-free-gp}.

Recently, significant efforts have been expanded to apply hierarchical matrices to induce low-rank block structure in the original dense matrices to improve both the storage complexity and the computation complexity for several common matrix operations, including matrix-vector products, factorizations and linear system solves. Various representations for hierarchical matrices have been proposed in the past \cite{hodlr-1,H-2-1,DH-2-1,hss-1,hierarchical-matrix-analysis} with active applications in different fields. The technique has also been proposed for Gaussian process computations and parameter estimations \cite{gp-using-h2,fast-gp-using-skeleton,gp-likelihood-using-h,fast-direct-method-for-gaussian-process,linear-cost-rf}, uncertainty quantification and conditional generation \cite{linear-inv-geo}, and high-dimensional Gaussian probability computations \cite{h-dimensional-gp-1,h-dimensional-gp-2}. In this work, we focus on a specific class of matrices referred to as as hierarchical off-diagonal low-rank (HODLR) matrices \cite{hodlr-1}. In our applications, the off-diagonal sub-blocks are well approximated by low-rank matrices and this structure significantly reduces the complexity of many linear algebra operations. In \cite{hodlr-1} it was shown that a typical HODLR matrix can be hierarchically factorized into a sequential product of block low-rank updates to the identity, yielding direct algorithms for linear system solve and evaluation of determinants in quasilinear scale. Both operations are essential components for evaluating the Gaussian likelihood. 

When a square matrix can be well-approximated by a hierarchical matrices structure, many methods can be applied to construct the hierarchical approximation via compressing the off-diagonal blocks to low-rank. In \cite{fast-direct-method-for-gaussian-process}, the authors provided  several ways to obtain the low-rank factorization of the off-diagonal blocks assuming the blocks are explicitly given. In \cite{h2-construct-1,h2-construct-2,hss-construct}, the methods for constructing hierarchical representations are presented relying on an $O(1)$ access to individual matrix entries. When explicit matrix entry access is not available, one can use randomized algorithms to efficiently approximate the target matrix from ``black box'' matrix-vector multiplication subroutines \cite{fast-construct-of-hierarchical-matrices,GPU-construction-of-hierarchical-matrices}.  Compared to the aforementioned explicit approaches, randomized methods avoid direct evaluations of the local blocks, which is essential when the dense matrix is costly to evaluate \cite{hierarchical-matrix-approx-of-hessian-in-inverse-problems}. The numerical accuracy of randomized algorithms has also been well-studied; see \cite{tropp-find-the-structure-with-randomness} for details.

In this work, we combine physics-based covariance models \cite{physics-based-cov-model} and randomized "black-box" hierarchical matrix construction \cite{hierarchical-matrix-approx-of-hessian-in-inverse-problems} and enhance them with derivative computations to design an efficient HODLR framework for Gaussian process maximum likelihood estimation. Compared to previous work, particularly \cite{physics-based-cov-model}, which proposed implicit covariance models assuming a low rank of the process covariance structure, and \cite{geoga2020scalable} which used HODLR for explicit Gaussian kernels, our contributions are that (a) we use an HODLR representation of implicitly defined kernels and propose an approach that leads to it being computed with only $O(\log n)$ forward model evaluations, (b) we use randomized sketching of the off-diagonal blocks and indicate a way to allow for the derivative information to be well approximated in the same structure, and (c) we propose an exact way of computing traces of products of HODLR matrices  which in turn allow the computation of the score function and Fischer information matrix for HODLR structures in $O(n \log^2 n)$.

The rest of the paper is structured as follows. In Section \ref{sec:cov model and MLE}, we review the physics-based covariance model and the maximum likelihood estimation procedure for parameter estimations. Section \ref{sec:HODLR approximation of cov matrices} contains a review of HODLR matrices techniques and linear algebra algorithms. We then introduce a matrix-free way, inspired by \cite{hierarchical-matrix-approx-of-hessian-in-inverse-problems} to construct the HODLR approximation via matrix-vector products of the targeting covariance matrix and differentiating the whole process to get its derivatives. In Section \ref{sec:HODLR approx of likelihood, scores, info matrix} we use the obtained hierarchical approximations to derive approximated log-likelihood, score equations and information matrix for parameter estimation. An application of the proposed algorithm to Gaussian wind field model is presented in Section \ref{sec:numerical experiments}. We end the paper with a discussion in Section \ref{sec:discussion}.

\section{Physics-based Covariance Models and Maximum Likelihood Estimation}
\label{sec:cov model and MLE}

To facilitate the description of our proposed approach, we assume a general physical model to work with. Consider a general deterministic physical model:
\begin{align}
    y=F(z),
    \label{eq:general physical model}
\end{align}
where $y$ is an $m$-dimensional random field and $z$ is an $n$-dimensional random field. $F:\mathbb{R}^m\rightarrow\mathbb{R}^n$ is a sufficiently regular mapping. Here $y$ is the output process (i.e. the process we can partially observe and want to predict) and $z$ is the latent process (i.e. the process that is not usually observable but is strongly correlated with the output process). $F$ can be interpreted as the physical relation that governs the processes. One example is the horizontal wind field model $U=\nabla\times\phi+\nabla\chi$ where $U$ is the horizontal wind component, $\phi$ is the stream function and $\chi$ is the velocity potential \cite{wind-model}. In terms of our model \eqref{eq:general physical model}, $y\xleftarrow{}U$ and $z\xleftarrow{}(\phi,\chi)$. This model will be used later in this study. 

We note that many physical models can be adapted or converted to this form. For example, many physical processes can be modeled using partial differential equations (PDEs). After proper discretizations the discretized PDE generally takes the form of \eqref{eq:general physical model}. One can also consider stochastic partial differential equations (SPDEs), if treating the random term as part of the latent process. Note that we will use the general form \eqref{eq:general physical model} in the following text. However in real applicable cases both the output and latent processes can be a concatenation of several independent processes. Separating independent variables and exploiting their correlations may introduce extra sparsity in the covariance model though we do not explore that avenue here. 

\subsection{Physics-based Covariance Models}
\label{sec:physics-based cov model}

In \cite{physics-based-cov-model}, the authors proposed a systematic way to incorporate the physical relation \eqref{eq:general physical model} into the design of covariance structure. Note that although \cite{physics-based-cov-model} only considers covariance models for spatial processes, the same approach can be adapted to spatio-temporal context with little extra effort.

Suppose we model the latent process as Gaussian processes. Specifically, let's assume $z=(z(x_1),z(x_2),\ldots,z(x_m))^T$, where $z(x)$ is a Gaussian random field indexed by a spatial location $x\in\mathbb{R}^d,d\geq 1$. Then $z$ follows a $N(\mu_z, \Sigma_z)$ distribution. Additionally the covariance matrix $\Sigma_z$ is further parameterized by a covariance function $K_{z}(\theta)$ which depends on a parameter vector $\theta\in\mathbb{R}^{p}$. Denote $\mathrm{E}(z)=\bar{z}$ and the perturbation around the mean by $\delta z=z - \bar{z}$. By \cite{physics-based-cov-model}, the covariance model of $y$ satisfies
\begin{align}
    K(\theta) = LK_z(\theta)L^T + O(||\delta z||^3),
    \label{eq:cov model}
\end{align}
where $L$ is the Jacobian matrix of $F$ evaluated at $\bar{z}$. Note that \eqref{eq:cov model} utilizes the linearization of $F$ which would incur a third-order error in $\delta z$. If the function $F$ is linear, the error term vanishes and \eqref{eq:cov model} becomes an exact covariance model. The covariance model can be applied when the underlying physical relation can be well approximated by its linearization. When the function is highly nonlinear, \cite{physics-based-cov-model} proposed a higher-order covariance model which can reduce the error to $O(||\delta z||^5)$ order, but due to its complexity we do not pursue that correction here and leave it to future work.

We note that in practice only a very limited fraction of the random field can be observed directly, i.e. $m\gg n$. So the covariance model \eqref{eq:cov model} is often of full-rank. Approximation methods based on the low-rank structure of the covariance matrix become very inaccurate in this case. Moreover, explicit construction of the covariance matrix becomes very costly since it requires matrix-matrix multiplications with large scale matrices. These facts prompt us to propose a low-rank approximation method for the off-diagonal blocks only in a hierarchical, fully matrix-free, manner that we discuss in \S \ref{sec:HODLR approximation of cov matrices}.

\subsection{Maximum Likelihood Estimation and Parameter Inference}
We carry out the estimation and inference by assuming that $y \sim \mathcal{N}(F(\bar{z}),K(\theta))$. If $F$ is a linear mapping the relationship is exact, otherwise, the distribution is only an approximation. We note, however, that the approach is quite common in nonlinear Bayesian inverse problems and extended Kalman filtering \cite{bayesian-parameter-estimation-with-uq,linearized-bayesian-inverse,kalman-filter-comparison,EKF-stability}. For the rest of the paper we will assume the distribution of $y$ to be the one stated, even if it may be only an approximation in practice. 
 Inferring the parameter vector $\theta=(\theta_1,\ldots,\theta_p)^T \in\mathbb{R}^p$ is of great scientific interest. Under our assumption about $y$ we get the (approximate, for $F$ nonlinear) log-likelihood function:
\begin{align}
    L(\theta)=-\frac{n}{2}\log(2\pi)-\frac{1}{2}\log{|K(\theta)|}-\frac{1}{2}(y-\bar{y})^T K(\theta)^{-1}(y-\bar{y}),
    \label{eq:log-like}
\end{align}
where $|A|$ denotes the determinant of square matrix $A$ and $\bar{y}=F(\bar{z})$. The maximum likelihood estimator of $\theta$ is the value $\hat{\theta}$ which maximizes \eqref{eq:log-like}. Since the nonzero mean 
$\bar{y}$ brings only simple algebraic changes in our algorithm, we will assume $\bar{y}$ to be $0$ in the rest of the text to simplify the notation. For the same reason, we drop the explicit dependence of $K$ on parameters $\theta$ in the rest of the paper. 

Evaluating the log-likelihood \eqref{eq:log-like} requires the evaluation of the log-determinant and the inverse of covariance matrix $K$ (or rather, the linear system solve of $K$). If the goal is to maximize the log-likelihood, alternatively one can avoid log-determinant computation by considering the score equations, which are obtained by setting the gradient of the log-likelihood function to be zero. The gradient of \eqref{eq:log-like} with respect to the parameters $\theta$ is given by
\begin{align}
    S_j(\theta)=-\frac{1}{2}\mathrm{tr}\left(K^{-1}K_j\right) + \frac{1}{2}y^T K^{-1}K_jK^{-1}y,\ j=1,\ldots,p,
    \label{eq:score}
\end{align}
where $S_j(\theta)$ denotes the derivative of $L$ with respect to parameter $\theta_j$, $K_j=\frac{\partial K}{\partial \theta_j}$ and $\mathrm{tr}(A)$ denotes the trace of square matrix $A$. One can find the minimizer of log-likelihood \eqref{eq:log-like} or equivalently the root of the score equations \eqref{eq:score} by quasi-Newton methods, for example, Broyden's method \cite{broyden-method}  \cite{practical-methods-of-optimization}, which only requires evaluations of the first-order score equations. However matrix-matrix operations with the inverse of $K$ are still required (even if carried one column at a time) to evaluate the term containing the trace.

Additionally, being able to efficiently estimate the observed Fisher information matrix can facilitate the uncertainty quantification of the given estimators, for example, building the confidence intervals of the estimates. Specifically, denote the maximum likelihood estimator (MLE) of the parameters by $\hat{\theta}$ and the observed Fisher information matrix evaluated at the maximum likelihood estimates by $\mathcal{I}(\hat{\theta})$. As the asymptotic theory suggests \cite{stein-theory-of-kriging}, if the smallest eigenvalue of $\mathcal{I}$ tends to infinity as the sample size grows, one can expect that
\begin{align} \label{eq:CLT}
    (\hat{\theta} - \theta^{\star})\xrightarrow{D} \mathcal{N} (0,\mathcal{I}(\hat{\theta})^{-1}),
\end{align}
where $\theta^\star$ represents the ground truth parameter values. We can then construct confidence intervals based on the asymptotic estimation.  Furthermore, for multivariate Gaussian processes where the uncertainties only occur in the covariance, each entry in the observed Fisher information matrix is given by
\begin{align} 
    \mathcal{I}_{i,j}(\hat{\theta}) = \frac{1}{2}\mathrm{tr}
    \left[{\left(K^{-1}K_iK^{-1}K_j\right)} |_{\theta=\hat{\theta}}\right].
    \label{eq:fisher info}
\end{align}
Even for circumstances where \eqref{eq:CLT} may not be guaranteed to hold, carrying out the maximum likelihood, and estimating the uncertainty in the parameters is very useful since it gives information whether the parameters are estimable in the first place, or whether only some of them may be \cite{crowder1976maximum}. This may be the case for the widely-used Matern covariance class, where only some parameters can be guaranteed to be estimable \cite{stein-theory-of-kriging}. As asymptotic normality relies on the accuracy of the Taylor expansion of the score equations at the true parameter \cite{crowder1976maximum}, a well behaved Fischer information matrix is an indication of approximate normality of the parameters being valid. 

Unfortunately evaluating \eqref{eq:log-like}, \eqref{eq:score} and \eqref{eq:fisher info} can be very costly. The matrix $K$  \eqref{eq:cov model} is generally dense (since $F(\cdot)$ may contain an inverse differential operator in cases of interest). A standard Cholesky factorization approach requires $O(n^3)$ computations and $O(n^2)$ memory since the covariance matrix is often unstructured and \eqref{eq:fisher info} requires matrix-matrix operations with the inverse of the covariance and not just solving linear systems with $K$.  

To circumvent these difficulties, in the next section we will utilize hierarchical matrix techniques, specifically HODLR matrices, to reduce both the computational and storage complexity to quasilinear scale. 

\section{HODLR Approximations for Covariance Matrices}
\label{sec:HODLR approximation of cov matrices}

In this section, we approximate the covariance matrix $K$ from \S \ref{sec:cov model and MLE} by the HODLR format \cite{hodlr-1} and define the algorithms that will compute the relevant components efficiently. 

\subsection{HODLR Matrices}
\label{sec:HODLR Matrices}

The HODLR matrix format uses a divide-and-conquer strategy to recursively divide the whole covariance matrix to sub-blocks and approximate all off-diagonal blocks by low-rank matrices. A typical 2-level symmetric positive definite matrix (SPD; the only kind relevant here due to properties of covariance matrices)-HODLR matrix $A\in\mathbb{R}^{n\times n}$ can be written in the following form:

\begin{align}
    A&=
    \begin{bmatrix}
    A_1^{(1)} & W_1^{(1)}X_1^{(1)T}\\
    V_1^{(1)}U_1^{(1)T} & A_2^{(1)}
    \end{bmatrix}\\
    &=
    \begin{bmatrix}
    \begin{bmatrix}
    A_1^{(2)} & W_1^{(2)}X_1^{(2)T}\\
    X_1^{(2)}W_1^{(2)T} & A_2^{(2)}
    \end{bmatrix} & W_1^{(1)}X_1^{(1)T}\\
    X_1^{(1)}W_1^{(1)T} & 
    \begin{bmatrix}
    A_3^{(2)} & W_2^{(2)}X_2^{(2)T}\\
    X_2^{(2)}W_2^{(2)T} & A_4^{(2)}
    \end{bmatrix}
    \end{bmatrix},
    \label{eq:hodlr format}
\end{align}
where the superscripts indicate the level of the approximation. All the off-diagonal blocks are subsequently approximated using low-rank matrices $W$, $X$ with the appropriate subscripts. For example in the first level, $W_1^{(1)},X_1^{(1)}\in\mathbb{R}^{n/2 \times k}$ where $k\leq n/2$. The local rank $k$ can be different for different levels (adaptive rank strategy) to guarantee the approximation error of each local block to be upper-bounded by some error tolerance $\epsilon$. Here for the simplicity of the presentation and complexity analysis we assume all local blocks have the same constant local rank $k$. For a 2-level HODLR matrix, the diagonal blocks $A_1^{(2)},A_2^{(2)},A_3^{(2)},A_4^{(2)}$ in the leaf level (level 2 in this case) are kept to remain full-rank (and identical to the respective sub blocks of the original matrix $A$). 

Taking advantage of the hierarchical structure, an SPD HODLR matrix admits several fast factorization algorithms, including its ``basic'' factorization introduced in \cite{fast-direct-method-for-gaussian-process}, symmetric factorization \cite{sym-factorization-of-hodlr}, QR factorization \cite{qr-factorization-of-hodlr}, LU factorization \cite{hm-toolbox}. Here we use the ``basic'' factorization to factorize the HODLR matrix to the product of a sequence of block low-rank updates of the identity matrix:

\begin{align}
    A = \bar{A}\prod_{i=\tau}^{1}(I+U^{(i)}V^{(i)T}),
    \label{eq:one way factor}
\end{align}
where $\bar{A}$ is a block-diagonal matrix containing all the leaf level diagonal blocks of $A$ (denoted by $A_i^{(2)}$ in \eqref{eq:hodlr format}), $I$ denotes the identity matrix of proper size and each $(I+U^{(i)}V^{(i)T})$ is a block-diagonal low-rank (rank-$2k$) update to the identity, for which the size of the diagonal blocks depends on the level $i$. Take $(I+U^{(2)}V^{(2)T})$ as an example, we have
\begin{align}
    (I+U^{(2)}V^{(2)T}) &= 
    \begin{bmatrix}
        I+U_1^{(2)}V_1^{(2)T} & 0\\
        0 & I+U_2^{(2)}V_2^{(2)T}
    \end{bmatrix},
    \label{eq:example-2nd level}
\end{align}
with two diagonal blocks. Generally, $(I+U^{(i)}V^{(i)T})$ has $2^{i-1}$ diagonal blocks with size $n/2^{i-1} \times n/2^{i-1}$. Each diagonal block is a rank $2k$ update to identity. For example in \eqref{eq:example-2nd level}, $U_1^{(2)},U_2^{(2)},V_1^{(2)},V_2^{(2)}$ are all of size $n/2^{i-1} \times 2k$. More details about the factorization can be found in Appendix \ref{app:hodlr-factorization}.

In \eqref{eq:one way factor}, $\tau$ is the number of levels of the HODLR matrix $A$. The factorization can be computed in $O(n\log^2{n})$ time if the local rank $k$ is fixed and the number of level grows with $O(\log{n})$. We note that an asymmetric HODLR matrix can also be factorized in the form of \eqref{eq:one way factor}, but symmetric factorization \cite{sym-factorization-of-hodlr} only exists for SPD HODLR matrices.

Utilizing the structure of the factorization \eqref{eq:one way factor}, the determinant of $A$ can be computed efficiently via Sylvester's determinant identity and the linear system can be 
efficiently solved by recursively applying the Woodbury matrix identity \cite{woodbury-identity}. To summarize, once the decomposition \eqref{eq:one way factor} is obtained, the subsequent numerical linear algebra operations including log-determinant computation, linear system solves and matrix-vector products can all be performed efficiently in at most $O(n\log^2{n})$ operations.  A list of common HODLR matrix linear algebra operations with their complexities are given in Table \ref{tb:hodlr operations}. For a more complete analysis of supported operations, we refer readers to \cite{hierarchical-matrix-analysis} for a discussion. 

\begin{table}[htb!]
\centering
\begin{tabular}{ |c|c|c|c|c|c|c|} 
 \hline
Ops. & \texttt{A*v} & \texttt{A\textbackslash v} & \texttt{det(A)} & \texttt{A+B} & \texttt{A*B} & \texttt{A\textbackslash B}\\
 \hline
Compl. & $O(n\log{n})$ & $O(n\log^2{n})$ & $O(n\log^2{n})$ & $O(n\log{n})$ & $O(n\log^2{n})$ & $O(n\log^2{n})$\\
 \hline
\end{tabular}
\caption{Complexity (Compl.) of common arithmetic operations (Ops.). $A,B$ are both HODLR matrices with size $n$, level $\log{n}$ and constant local rank. $v$ is a vector of length $n$.}
\label{tb:hodlr operations}
\end{table}

We note that all operations involving two HODLR matrices in Table \ref{tb:hodlr operations} will lead to increase of off-diagonal ranks. The computational efficiency of the resulting matrix will significantly decrease due to the accumulation of low-rank updates from recursive calls. To tackle the issue, one can combine the arithmetic operations with re-compression to reduce the rank after the computation of each level.  Most studies conduct these operations inexactly and heavily rely on the assumptions that the off-diagonal ranks encountered during an operation remain $O(k)$. In this study, we will also encounter operations involving two HODLR matrices. Based on the specific needs of our computation, we derive two new types of HODLR operations which can be conducted exactly and in quasilinear time scale. Details can be found in Section \ref{new operations for hodlr}.

In terms of storage complexity, both the HODLR matrix and its factors take $O(n\log{n})$ memory. The remaining problem is to construct the HODLR approximation for the given covariance matrix \eqref{eq:cov model}.

\subsection{Randomized Matrix-free Construction of HODLR Matrices}

Assume the  SPD matrix $K=LK_zL^T$ in \eqref{eq:cov model} can be well-approximated by the HODLR format. Note that in our covariance model \eqref{eq:cov model}, $m\gg n$ thus the matrix $K_z$ may be difficult to store. Moreover, the matrix $L$ may be exceedingly difficult to obtain explicitly since the direct model $F$ may be arbitrarily complex, for example a climate code. Therefore, if the task is to construct the HODLR approximation of $K$, the direct approach dividing the dense covariance matrix recursively and constructing the low-rank approximation for each off-diagonal block (for example, using SVD) is infeasible. 

We then turn to the feasibility of computing the HODLR approximation of $K$ by accessing it only by means of matrix-vector products $K \tilde{y}$, which in turn requires efficient access to matrix-vector products with $L$, $K_z$, and $L^T$. Since $L$ is the Jacobian of the operator $F$, each $L$-vector and $L^T$-vector product can be evaluated via forward mode and reverse mode automatic differentiation (AD) \cite{AD} with the same order of cost as evaluating a forward solve of the physical model $F$. The covariance matrix $K_z$ can be structured (e.g. HODLR itself) or sparse,  and thus it can be reasonably assumed to allow for fast matrix-vector products. 
Therefore in many circumstances, $L$ and $K_z$ admits fast and efficient matrix-vector multiplication operations. We refer readers to Section \ref{sec:numerical experiments} for examples. 

We now turn to the issue of building the HODLR approximation of $K$.  If one explicitly constructs the covariance matrix $K$ first, this generally takes $O(m)$ forward model evaluations and storing the covariance matrix takes $O(n^2)$ memory, which may be infeasible. An important task, consequently, will be to enable the computation of the HODLR approximation of $K$ in \eqref{eq:cov model} using much less memory and forward model evaluations. 
For the rest of the paper, we assume there exists a fast solver of computing matrix-vector products for covariance matrix $K$ with any vector and we track the number of $K$-vector products for complexity analysis, aiming to get it much smaller than the brute force $O(m)$.


To this end, randomized algorithms for matrix factorizations have proven to be robust and accurate \cite{tropp-find-the-structure-with-randomness}.  Here we adapt the idea in \cite{fast-construct-of-hierarchical-matrices} to HODLR matrices and propose a way to construct the HODLR approximation of the covariance matrix $K$ using only $K$-vector multiplications in \S \ref{ss:rmf} and \S \ref{sec:hodlr approx of cov matrix} in conjunction with randomized algorithms to efficiently produce the low-rank approximations to the off-diagonal blocks.

\subsubsection{Randomized Matrix factorization}\label{ss:rmf}
 We adopt the simplest randomized matrix factorization algorithm \cite{tropp-find-the-structure-with-randomness} to facilitate our computation and presentation, though other variants of the algorithm including randomized SVD \cite{random-svd} can also be used here with little modification. For a given matrix $B$ and a given target rank $k$, the algorithm is summarized in Algorithm \ref{alg:randomized factorization}.

\begin{algorithm}[!htb]
\caption{Randomized Low-rank Approximation}\label{alg:randomized factorization}
\textbf{Input}: A matrix $B\in\mathbb{R}^{p\times q}$ and a given target rank $k$, $k<q$.

\textit{Step 1}. Draw a Gaussian random sampling matrix $\Omega\in\mathbb{R}^{p\times k}$.

\textit{Step 2}. Compute $Y=B\Omega$. Now $Y\in\mathbb{R}^{p\times k}$.

\textit{Step 3}. Compute the QR factorization for $Y$ with column pivoting. Denote the Q factor by $Q\in\mathbb{R}^{p\times k}$.

\textit{Step 4}. Compute $Q^T B$ by $(B^T Q)^T$.

\textbf{Output}: $\hat{B}= Q(Q^T B)$ is a rank-$k$ approximation of $B$.
\end{algorithm}

The accuracy of the algorithm and its variants has been well-studied \cite{tropp-find-the-structure-with-randomness,random-matrix-approx-1,random-svd}. When using the matrix 2-norm to measure the approximation error, the minimal error is $\sigma_{k+1}$ which is the $(k+1)$-th largest singular value and the lower bound is achieved by the singular vector decomposition (SVD) truncated to the rank $k$. It can be shown that the approximation error can be bounded by $k,c$ and $\sigma_k$. Specifically, when the singular values of $B$ decay rapidly, Algorithm \ref{alg:randomized factorization} can provide an approximation that is very close to the optimal truncated SVD solution with high probability \cite{random-matrix-approx-1}.
\subsubsection{Construction From Matrix-vector Products}
\label{sec:hodlr approx of cov matrix}

To discuss our construction of the HODLR matrix approximation \eqref{eq:hodlr format}, using Algorithm \ref{alg:randomized factorization} and employing only $O(\log_2(n)k)$  matrix-vector multiplications, we follow the approach in \cite{martinsson2016compressing}. The idea is to compress the off-diagonal blocks via randomized sampling. In this subsection we review the HODLR compression procedure to lay the foundation for the next several subsections. Note that our approach goes beyond \cite{martinsson2016compressing} by further considering the derivatives of the construction procedure.

Assume we work with a two-level  covariance  matrix $K$:
\begin{align}
    K = \begin{bmatrix}
    \begin{bmatrix}
    A_1^{(2)} & B_1^{(2)}\\
    B_1^{(2)T} & A_2^{(2)}
    \end{bmatrix} & B_1^{(1)}\\
    B_1^{(1)T} & 
    \begin{bmatrix}
    A_3^{(2)} & B_2^{(2)}\\
    B_2^{(2)T} & A_4^{(2)}
    \end{bmatrix}
    \end{bmatrix}.
    \label{eq:full K in HODLR format}
\end{align}


While the algorithm does not have requirements about the matrix size, we will assume $n$ is divisible by $4$ for simplifying the presentation.

\textit{Processing level $1$}. Assume the row index set and the column index set of $B_1^{(1)}$ are given by $\mathcal{I}_1^{(1)}$ $\left(\stackrel{\Delta}{=}1:\frac{n}{2}\right)$ and $\mathcal{I}_2^{(1)}$ $\left(\stackrel{\Delta}{=}\left(\frac{n}{2}+1 \right):n\right)$ respectively. We draw a $(n/2)\times k$ sampling matrix $R_1^{(1)}$ with i.i.d standard normal entries. Then we construct a patterned $n\times k$ sampling matrix by filling $R^{(1)}(\mathcal{I}_1^{(1)},:)$ with 0 and $R^{(1)}(\mathcal{I}_2^{(1)},:)=R_1^{(1)}$. Here we use the MATLAB notation $R^{(1)}(\mathcal{I}_1^{(1)},:)$ to indicate the selected rows from $R^{(1)}$ according to the index set $\mathcal{I}_1^{(1)}$. Now we observe that $R^{(1)^T}=\left[ 0 \quad R_1^{(1)^T}\right]$ and thus 
\begin{align}
    K R^{(1)}=
    K
    \begin{bmatrix}
    0\\R_1^{(1)}
    \end{bmatrix}=
    \begin{bmatrix}
    B_1^{(1)}R_1^{(1)}\\
    \begin{bmatrix}
    A_3^{(2)} & B_2^{(2)}\\
    B_2^{(2)T} & A_4^{(2)}
    \end{bmatrix}R_1^{(1)}
    \end{bmatrix}.
    \label{eq:lvl1 sample1}
\end{align}
By restricting the row index of the right-hand side to index set $\mathcal{I}_1^{(1)}$ and discarding the bottom half, we get $B_1^{(1)}R_1^{(1)}$ as the randomly sampled column space of $B_1^{(1)}$. The result corresponds to steps 1 and 2 in Algorithm \ref{alg:randomized factorization}. For carrying out step 3, we compute the QR factorization of $B_1^{(1)}R_1^{(1)}$. Denote the orthogonal basis matrix by $Q_1^{(1)} \in \mathbb{R}^{\frac{n}{2} \times k}$. To compute $Q_1^{(1)T}B_1^{(1)}$ we form another patterned matrix $S^{(1)}\in\mathbb{R}^{n\times k}$ by filling $S^{(1)}(\mathcal{I}_2^{(1)},:)$ with 0 and $S^{(1)}(\mathcal{I}_1^{(1)},:)=Q_1^{(1)}$. Note that $S^{(1)^T}=\left[ Q_1^{(1)^T} 0 \right]$ and 
\begin{align}
    K S^{(1)}=
    K
    \begin{bmatrix}
    Q_1^{(1)}\\0
    \end{bmatrix}=
    \begin{bmatrix}
    \begin{bmatrix}
    A_1^{(2)} & B_1^{(2)}\\
    B_1^{(2)T} & A_2^{(2)}
    \end{bmatrix}Q_1^{(1)}\\
    B_1^{(1)T}Q_1^{(1)}
    \end{bmatrix}.
    \label{eq:lvl1 sample2}
\end{align}

Likewise, we restrict the output to the row index set $\mathcal{I}_2^{(1)}$ and discard the first half to obtain $\tilde{B}_1^{(1)}\stackrel{\Delta}{=}B_1^{(1)T}Q_1^{(1)}$. We then obtain \textit{the low-rank (rank $k$) representation of the off-diagonal block}
\begin{equation} \label{eq:lowrnkB11}
	\hat{B}_1^{(1)}= Q_1^{(1)} \tilde{B}_1^{(1)^T}= Q_1^{(1)}(Q_1^{(1)T}B_1^{(1)}).
\end{equation}	

\textit{Processing level $2$}. Next we move to the blocks of the second level in the hierarchy, i.e. $B_1^{(2)}$  and $B_2^{(2)}$. We draw a patterned sampling matrix of  $R^{(2)}=(0,R_1^{(2)T},0,R_2^{(2)T})^T\in\mathbb{R}^{n\times k}$, with $R_i^{(2)} \in \mathbb{R}^{\frac{n}{4} \times k}$, $i=1,2$, having Gaussian entries.  Here, the nonzero rows of $R^{(2)}$ correspond to column indices of $B_1^{(2)}$ $\left( \stackrel{\Delta}{=}  \left(\frac{n}{4}+1\right):\frac{n}{2} \right)$ and $B_2^{(2)}$ $\left( \stackrel{\Delta}{=}  \left(\frac{3n}{4}+1\right):n \right)$ 
 in $K$. We observe that
\begin{align}
    \left(K -
    \begin{bmatrix}
    0&\hat{B}_1^{(1)}\\
    \hat{B}_1^{(1)T}&0
    \end{bmatrix}\right)
    \begin{bmatrix}
    0\\R_1^{(2)}\\0\\R_2^{(2)}
    \end{bmatrix}
    \approx
    \begin{bmatrix}
    	\begin{bmatrix}
    		A_1^{(2)} & B_1^{(2)}\\
    		B_1^{(2)T} & A_2^{(2)}
    	\end{bmatrix} & 0\\
    	0 & 
    	\begin{bmatrix}
    		A_3^{(2)} & B_2^{(2)}\\
    		B_2^{(2)T} & A_4^{(2)}
    	\end{bmatrix}
    \end{bmatrix}
    \begin{bmatrix}
	0\\R_1^{(2)}\\0\\R_2^{(2)}
\end{bmatrix}
=
    \begin{bmatrix}
    B_1^{(2)}R_1^{(2)}\\
    A_2^{(2)}R_1^{(2)}\\
    B_2^{(2)}R_2^{(2)}\\
    A_4^{(2)}R_2^{(2)}
    \end{bmatrix}.
    \label{eq:lvl2 sample1}
\end{align}

Note that it is impossible to directly calculate the matrix subtraction on the left-hand side of \eqref{eq:lvl2 sample1} since the matrix $K$ is not explicitly available. Instead, we access $K$ only via matrix-vector products. Furthermore, we only store the obtained low-rank factors of $\hat{B}_1^{(1)}$ explicitly. The $\hat{B}_1^{(1)}$-vector or $\hat{B}_1^{(1)}$-matrix products are done by sequentially applying both low-rank factors ($Q_1^{(1)}$ and $\tilde{B}_1^{(1)^T}$) to the target matrix or vector. 


We form the new patterned matrix  $S^{(2)}=(Q_1^{(2)T},0,Q_2^{(2)T},0)^T\in\mathbb{R}^{n\times k}$, where the positioning is such that the zero blocks are square. Then we can compute
\begin{align}
    \left(K -
    \begin{bmatrix}
    0&\hat{B}_1^{(1)}\\
    \hat{B}_1^{(1)T}&0
    \end{bmatrix}\right)
    \begin{bmatrix}
    Q_1^{(2)}\\0\\Q_2^{(2)}\\0
    \end{bmatrix}
  \approx
\begin{bmatrix}
	\begin{bmatrix}
		A_1^{(2)} & B_1^{(2)}\\
		B_1^{(2)T} & A_2^{(2)}
	\end{bmatrix} & 0\\
	0 & 
	\begin{bmatrix}
		A_3^{(2)} & B_2^{(2)}\\
		B_2^{(2)T} & A_4^{(2)}
	\end{bmatrix}
\end{bmatrix}
    \begin{bmatrix}
	Q_1^{(2)}\\0\\Q_2^{(2)}\\0
\end{bmatrix}
=
    \begin{bmatrix}
    A_1^{(2)}Q_1^{(2)}\\
    B_1^{(2)T}Q_1^{(2)}\\
    A_3^{(2)}Q_2^{(2)}\\
    B_2^{(2)T}Q_2^{(2)}
    \end{bmatrix}.
    \label{eq:lvl2 sample2}
\end{align}


Similar to the first step, we now extract the even block entries from the right-hand side of \eqref{eq:lvl2 sample2}, that is the ones corresponding to rows indices $\left(\frac{n}{4}+1\right):\frac{n}{2}$, $\tilde{B}_1^{(2)}=  B_1^{(2)T}Q_1^{(2)}$, and $\left(\frac{3n}{4}+1\right):n$,  $\tilde{B}_2^{(2)}=  B_2^{(2)^T}Q_2^{(2)}$. 
We then obtain \textit{the low rank representation of the off-diagonal blocks on the second level} 
\begin{equation}\label{eq:lowrnkB212}
\hat{B}_1^{(2)}=Q_1^{(2)} \tilde{B}_1^{(2)^T}=Q_1^{(2)}(Q_1^{(2)T}B_1^{(2)}), \quad \hat{B}_2^{(2)}=Q_2^{(2)} \tilde{B}_2^{(2)^T}=Q_2^{(2)}(Q_2^{(2)T}B_2^{(2)}). 
\end{equation}
If there are finer levels, we can proceed to the next level in a similar fashion by first ``removing'' all off-diagonal blocks from the previous levels. 

\textit{Processing the leaf level}. To sketch the leaf level $A_1^{(2)},A_2^{(2)},A_3^{(2)},A_4^{(2)}$, we construct sampling matrix $S$ of size $n\times \mathrm{size(A_1^{(2)})}$ which is a vertical concatenation of identity matrices, $S=\left[ I_{\frac{n}{4}}, I_{\frac{n}{4}}, I_{\frac{n}{4}}, I_{\frac{n}{4}} \right]^T$. We will exploit the fact that we have an approximation of all level $1$ and $2$ off-diagonal blocks,
\begin{align}
    \left(K - \begin{bmatrix}
    \begin{bmatrix}
    0 & \hat{B}_1^{(2)}\\
    \hat{B}_1^{(2)T} & 0
    \end{bmatrix} & \hat{B}_1^{(1)}\\
    \hat{B}_1^{(1)T} & 
    \begin{bmatrix}
    0 & \hat{B}_2^{(2)}\\
    \hat{B}_2^{(2)T} & 0
    \end{bmatrix}
    \end{bmatrix}\right)S \approx
\begin{bmatrix}
	\begin{bmatrix}
		A_1^{(2)} & 0 \\
		0 & A_2^{(2)}
	\end{bmatrix} & 0\\
	0 & 
	\begin{bmatrix}
		A_3^{(2)} & 0\\
		0 & A_4^{(2)}
	\end{bmatrix}
\end{bmatrix}
    \begin{bmatrix}
	 I_{\frac{n}{4}} \\  I_{\frac{n}{4}}  \\  I_{\frac{n}{4}} \\  I_{\frac{n}{4}} 
\end{bmatrix}
=
    \begin{bmatrix}
        A_1^{(2)}\\
        A_2^{(2)}\\
        A_3^{(2)}\\
        A_4^{(2)}
    \end{bmatrix}.
    \label{eq:leaf sample}
\end{align}

\textit{Asymptotic complexity}. Assume all off-diagonal blocks have the same rank $k$ and the number of levels is $\tau=log_2\left(\frac{n}{n_l}\right)$, where $n_l$ is the size of the leaf block. Overall, the entire procedure requires $O(k\tau)$ $K$-vector products and $O(nk^2\tau^2)$ time complexity. If we assume constant off-diagonal rank $k=O(1)$ and the number of levels grows as $O(\log{n})$, for example $\tau=\lfloor\log_2{(\frac{n}{k})}\rfloor$, the computational complexity is $O(\log{n})$ $K$-vector products and $O(n\log^2{n})$ complexity. The storage complexity is $O(n\log{n})$. More details about the complexity analysis can be found in \cite[Section 4.1]{martinsson2016compressing}. These assumptions are consistent with the assumptions in \cite{sym-factorization-of-hodlr, martinsson2016compressing} for complexity analysis.

\subsection{Differentiating the HODLR Approximation}

With the HODLR approximation of the covariance matrix $K$ at hand, we can now handle the log-determinant term and the matrix inverse term of $K$ in the log-likelihood \eqref{eq:log-like} and score \eqref{eq:score} efficiently. However we now need to compute or  produce an adequate approximation of $K_j$. To this end, we will also approximate the derivatives of the covariance matrix in HODLR format. We will explore the fact that our HODLR approximation consists of a sequence of covariance matrix-vector products, block subsettings, and QR factorizations. The entire process is differentiable, as long as the patterned sampling matrices (e.g. $S^{(1)}$, $S^{(2)}$, $R^{(1)}$, $R^{(2)} $ in \S \ref{sec:hodlr approx of cov matrix} ) and subsetting patterns are kept fixed. We thus differentiate the approximation algorithm while simultaneously constructing the HODLR approximation for both the covariance matrix and its derivatives. 

As an illustrative example, we follow the same framework as in \S \ref{sec:hodlr approx of cov matrix} and approximate $K$ and its derivatives to 2-level HODLR form. 

\textit{Processing level $1$}. Starting from the first level we  differentiate \eqref{eq:lvl1 sample1} with respect to any parameter $\theta_j$ using forward mode AD \cite{AD} with the same order of cost as evaluating the $K$-vector product. Recall that we keep $R^{1}$ fixed, that is independent of $\theta$. In other words we draw $R^{(1)}$ once and keep it fixed for all optimization iterations. Differentiating through the subsetting operation (the top matrix on level 1 in \S \ref{sec:hodlr approx of cov matrix} ) we obtain the identity
\begin{align}\label{eq:db1r1}
     \frac{\partial B_1^{(1)} R^{(1)}}{\partial \theta_j}=\frac{\partial K R^{(1)}}{\partial \theta_j}(\mathcal{I}_1^{(1)},:).
\end{align}
In the next step we differentiate the QR factorization of $B_1^{(1)}R_1^{(1)}$, specifically the Q factor $Q_1^{(1)}$ as summarized in Algorithm \ref{alg:differentiate QR}, \cite{perturbation-of-qr}, to obtain $\frac{\partial Q_1^{(1)}}{\partial \theta_j}$.

\begin{algorithm}[!htb]
\caption{Differentiate the QR factorization \cite{perturbation-of-qr}}\label{alg:differentiate QR}
\textbf{Input}: A full column rank matrix $B\in\mathbb{R}^{p\times q}$ where $p>q$. Assume the (compact) QR factorization of $B$ is given by 
\begin{align}
    B = Q_1 R, \quad Q_1 \in \mathbb{R}^{p \times q}, \; Q_1^T Q_1 = I_q,\; R \in \mathbb{R}^{q \times q}
\end{align}
Also assume $dB$ is known, the differentiation can be computed in the following steps.

\textit{Step 1}. Form $Y=Q_1^T dB R^{-1}$.

\textit{Step 2}. Let $d\Omega$ be a $q\times q$ skew-symmetric matrix. Notice that both $R$ and its differentiation $dR$ are upper triangular matrices. Therefore $d\Omega$ and $dR$ can be uniquely computed from the following identity (derived from the product rule $dB=dQ_1R+Q_1dR$ inserted in $Y$)
\begin{align}
    Y=d\Omega + dR R^{-1}.
\end{align}

\textit{Step 3}. Compute 
\begin{align}
    dQ_1 &= Q_1 d\Omega + Q_2 Q_2^T dB R^{-1}
    \label{eq:diff qr original}
    \\
    &= Q_1 d\Omega + (I - Q_1 Q_1^T) dB R^{-1}.
    \label{eq:diff qr adapted}
\end{align}

\textbf{Output}: $dQ_1$, $dR$.
\end{algorithm}

At the next step of the Algorithm from \S \ref{sec:HODLR approximation of cov matrices}, we compute $B_1^{(1)T}Q_1^{(1)}$ via \eqref{eq:lvl1 sample2}. By differentiating the $K$-vector product and keeping in mind the dependence between $Q_1^{(1)}$ and parameters $\theta_j$, we can obtain $\frac{\partial B_1^{(1)T} Q_1^{(1)}}{\partial \theta_j}$,

\begin{align}
    \frac{\partial B_1^{(1)T} Q_1^{(1)}}{\partial \theta_j} &= \left(\frac{\partial B_1^{(1)}}{\partial \theta_j}\right)^T Q_1^{(1)} + B_1^{(1)T} \frac{\partial Q_1^{(1)}}{\partial \theta_j}, \label{eq:prod-rule-1}\\
    &\stackrel{\eqref{eq:lvl1 sample2},\eqref{eq:full K in HODLR format}}{=} \frac{\partial K S^{(1)}}{\partial \theta_j}(\mathcal{I}_2^{(1)},:) + K \begin{bmatrix}
    \frac{\partial Q_1^{(1)}}{\partial \theta_j}\\0
    \end{bmatrix}(\mathcal{I}_2^{(1)},:). \label{eq:prod-rule-2}
\end{align}

Note that we cannot compute \eqref{eq:prod-rule-1} directly since the kernel matrix $K$ is not explicitly available. Instead we use the same trick as in \eqref{eq:lvl1 sample2} and rewrite both terms with $K$-vector products as in \eqref{eq:prod-rule-2}. Specifically we can compute the first term in \eqref{eq:prod-rule-2} via AD assuming the factor $Q_1^{(1)}$ is held as a constant, as is the case for $S^{(1)}$ in \eqref{eq:lvl1 sample2}. The second term can be computed by $k$ $K$-vector products with derivatives $\frac{\partial Q_1^{(1)}}{\partial \theta_j}$.

Recall we approximate $B_1^{(1)}$ by low-rank factorization $\hat{B}_1^{(1)}= Q_1^{(1)}(Q_1^{(1)T}B_1^{(1)})$, \eqref{eq:lowrnkB11}. We differentiate both sides by parameter $\theta_j$ to obtain \textit{the low-rank representation of the derivative of the first-level off-diagonal block, $B_1^{(1)}$:}

\begin{align}
    \frac{\partial \hat{B}_1^{(1)}}{\partial \theta_j}=&\frac{\partial Q_1^{(1)}}{\partial \theta_j} \left(B_1^{(1)T}Q_1^{(1)}\right)^T + Q_1^{(1)} \left(\frac{\partial B_1^{(1)T} Q_1^{(1)}}{\partial \theta_j}\right)^T.
    \label{eq:diff low-rank}
\end{align}

We now summarize the workflow that we used to efficiently compute the derivative of the low approximation since we will apply this philosophy at each level:
	\begin{enumerate}
		\item We differentiate \eqref{eq:lvl1 sample1}, we get $\frac{\partial B_1^{(1)}R_1^{(1)}}{\partial \theta_j}$, \eqref{eq:db1r1}, using differentiation of $k$ $K$-vector products.
		\item  We use Algorithm  \ref{alg:differentiate QR} to differentiate the QR factorization of  $B_1^{(1)}R_1^{(1)}$. We obtain $\frac{\partial Q_1^{(1)}}{\partial \theta_j}$. 
		\item We differentiate \eqref{eq:lvl1 sample2}, exploit the HODLR structure  \eqref{eq:full K in HODLR format} and obtain the first term in \eqref{eq:prod-rule-2}.This steps requires differentiation of $k$ $K$-vector products. Then 
		we use another $k$ $K$-vector products with the derivatives of the $Q$-factors to form the second term of \eqref{eq:prod-rule-2}. Combining two terms together, we obtain $\frac{\partial B_1^{(1)T} Q_1^{(1)}}{\partial \theta_j}$.
		\item The derivative of the low-rank approximation, $\frac{\partial \hat{B}_1^{(1)}}{\partial \theta_j}$, is available by means of the right hand side of \eqref{eq:diff low-rank}.
	\end{enumerate}
Note  that both terms in \eqref{eq:diff low-rank} are of rank $k$, which means the rank of $\frac{\partial \hat{B}_1^{(1)}}{\partial \theta_j}$ is at most $2k$. In practice we never construct \eqref{eq:diff low-rank} explicitly. Instead we store the low-rank components of both terms in \eqref{eq:diff low-rank} and invoke them when necessary. Also note that our workflow requires $3k$ matrix-vector products with either the covariance matrix $K$ or the covariance matrix derivative $K_j$ and $nk$ storage for the components (two $\frac{n}{2} \times k$ blocks).

\textit{Processing level $2$}. The second-level  procedure is the same as for the first level except the approximations from level 1 need to be removed.  We start by differentiating \eqref{eq:lvl2 sample1},

\begin{align}
    \left(K_j -
    \begin{bmatrix}
    0&\frac{\partial \hat{B}_1^{(1)}}{\partial \theta_j}\\
    \frac{\partial \hat{B}_1^{(1)T}}{\partial \theta_j}&0
    \end{bmatrix}\right)
    \begin{bmatrix}
    0\\R_1^{(2)}\\0\\R_2^{(2)}
    \end{bmatrix}
    =
    \frac{\partial}{\partial \theta_j} \left(K\begin{bmatrix}
    0\\R_1^{(2)}\\0\\R_2^{(2)}
    \end{bmatrix}\right) - 
    \begin{bmatrix}
    \frac{\partial \hat{B}_1^{(1)\ }}{\partial \theta_j}\begin{bmatrix}
    0\\R_2^{(2)}
    \end{bmatrix}\\
    \frac{\partial \hat{B}_1^{(1)T}}{\partial \theta_j}\begin{bmatrix}
    0\\R_1^{(2)}
    \end{bmatrix}
    \end{bmatrix}
    \approx
    \begin{bmatrix}
    \frac{\partial B_1^{(2)}R_1^{(2)}}{\partial \theta_j}\\
    \frac{\partial A_2^{(2)}R_1^{(2)}}{\partial \theta_j}\\
    \frac{\partial B_2^{(2)}R_2^{(2)}}{\partial \theta_j}\\
    \frac{\partial A_4^{(2)}R_2^{(2)}}{\partial \theta_j}
    \end{bmatrix}.
    \label{eq:lvl2 sample1 diff}
\end{align}

Note that we do not explicitly differentiate the matrices on the left-hand side of \eqref{eq:lvl2 sample1 diff} since we do not have direct access to matrix $K$. Instead we differentiate the matrix-vector products in \eqref{eq:lvl2 sample1} (see the expression in the middle). The first term in the middle expression is computed via AD by differentiating the $K$-vector products. The second term is computed by matrix-vector products with the low-rank representation of $\frac{\partial \hat{B}_1^{(1)}}{\partial \theta_j}$ \eqref{eq:diff low-rank}. Now we obtain $\frac{\partial B_1^{(2)}R_1^{(2)}}{\partial \theta_j}$ and $\frac{\partial B_2^{(2)}R_1^{(2)}}{\partial \theta_j}$ by truncating the right-hand side of \eqref{eq:lvl2 sample1 diff}. We use these low-rank matrices in Algorithm \ref{alg:differentiate QR} and obtain the derivatives of the $Q$-factors with respect to the parameters: $\frac{\partial Q_1^{(2)}}{\partial \theta_j}$, $\frac{\partial Q_2^{(2)}}{\partial \theta_j}$.

Next, we differentiate \eqref{eq:lvl2 sample2} (reversing the order of terms in that equation to make our argument) by treating $Q_1^{(2)}$, $Q_2^{(2)}$, components of $S^{(2)}$, as constant matrices, 

\begin{align}
	    \begin{bmatrix}
		\frac{\partial A_1^{(2)}}{\partial \theta_j}Q_1^{(2)}\\
		\frac{\partial B_1^{(2)T}}{\partial \theta_j}Q_1^{(2)}\\
		\frac{\partial A_3^{(2)}}{\partial \theta_j}Q_2^{(2)}\\
		\frac{\partial B_2^{(2)T}}{\partial \theta_j}Q_2^{(2)}
	\end{bmatrix}
\stackrel{\eqref{eq:lvl2 sample2}, \eqref{eq:full K in HODLR format}}{\approx}
    \left(K_j -
    \begin{bmatrix}
    0&\frac{\partial \hat{B}_1^{(1)}}{\partial \theta_j}\\
    \frac{\partial \hat{B}_1^{(1)T}}{\partial \theta_j}&0
    \end{bmatrix}\right)
    \begin{bmatrix}
    Q_1^{(2)}\\0\\Q_2^{(2)}\\0
    \end{bmatrix}
    =
    \frac{\partial}{\partial \theta_j}\left(K\begin{bmatrix}
    Q_1^{(2)}\\0\\Q_2^{(2)}\\0
    \end{bmatrix}\right) 
- \begin{bmatrix}
    \frac{\partial \hat{B}_1^{(1)T\ }}{\partial \theta_j}\begin{bmatrix}
    Q_2^{(2)}\\0
    \end{bmatrix}\\
    \frac{\partial \hat{B}_1^{(1)}}{\partial \theta_j}\begin{bmatrix}
    Q_1^{(2)}\\0
    \end{bmatrix}
    \end{bmatrix}.
    \label{eq:lvl2 sample2 diff}
\end{align}
In the last term, the first component only is the one requiring access to $K$ and the real computation weight. It can be computed via AD by differentiating $K$-vector products and treating $Q_1^{(2)}$, $Q_2^{(2)}$ as constant matrices. For the second component of the last term we do matrix-vector products with the low-rank representation of $\frac{\partial \hat{B}_1^{(1)}}{\partial \theta_j}$ \eqref{eq:diff low-rank} . Furthermore, by replacing $Q_1^{(2)}$, $Q_2^{(2)}$ in \eqref{eq:lvl2 sample2} with $\frac{\partial Q_1^{(2)}}{\partial \theta_j}$, $\frac{\partial Q_2^{(2)}}{\partial \theta_j}$, reverting its order, and using \eqref{eq:full K in HODLR format} we have

\begin{align}
	    \begin{bmatrix}
		A_1^{(2)}\frac{\partial Q_1^{(2)}}{\partial \theta_j}\\
		B_1^{(2)T}\frac{\partial Q_1^{(2)}}{\partial \theta_j}\\
		A_3^{(2)}\frac{\partial Q_2^{(2)}}{\partial \theta_j}\\
		B_2^{(2)T}\frac{\partial Q_2^{(2)}}{\partial \theta_j}
	\end{bmatrix}
\approx
    \left(K -
    \begin{bmatrix}
    0&\hat{B}_1^{(1)}\\
    \hat{B}_1^{(1)T}&0
    \end{bmatrix}\right)
    \begin{bmatrix}
    \frac{\partial Q_1^{(2)}}{\partial \theta_j}\\0\\\frac{\partial Q_2^{(2)}}{\partial \theta_j}\\0
    \end{bmatrix}.
    \label{eq:tmp2}
\end{align}
Now we subset the left-hand side of \eqref{eq:lvl2 sample2 diff} and \eqref{eq:tmp2} to the terms involving $B_1^{(2)}, B_2^{(2)}$ and use the product rule to compute

\begin{align}
   \label{eq:derbq12}  \frac{\partial (B_1^{(2)T} Q_1^{(2)})}{\partial \theta_j} = \frac{\partial B_1^{(2)T}}{\partial \theta_j}Q_1^{(2)} + B_1^{(2)T} \frac{\partial Q_1^{(2)}}{\partial \theta_j},\\
   \label{eq:derbq22} \frac{\partial (B_2^{(2)T} Q_2^{(2)})}{\partial \theta_j} = \frac{\partial B_2^{(2)T}}{\partial \theta_j}Q_2^{(2)} + B_2^{(2)T} \frac{\partial Q_2^{(2)}}{\partial \theta_j}.
\end{align}

We apply the product rule to \eqref{eq:lowrnkB212} to obtain \textit{the low-rank representations of the derivatives of the second level off-diagonal blocks:}

\begin{align} 
    &\frac{\partial \hat{B}_1^{(2)}}{\partial \theta_j}
    =\frac{\partial Q_1^{(2)}}{\partial \theta_j} \left(B_1^{(2)T}Q_1^{(2)}\right)^T + Q_1^{(2)} \left(  \frac{\partial (B_1^{(2)T} Q_1^{(2)})}{\partial \theta_j} \right)^T, 
    \label{eq:lvl2 block1 diff}\\
    &\frac{\partial \hat{B}_2^{(2)}}{\partial \theta_j}
    =\frac{\partial Q_2^{(2)}}{\partial \theta_j} \left(B_2^{(2)T}Q_2^{(2)}\right)^T + Q_2^{(2)}\left( \frac{\partial (B_2^{(2)T} Q_2^{(2)})}{\partial \theta_j} \right)^T.
    \label{eq:lvl2 block2 diff}
\end{align}

To access this representation, we store the low-rank components in the right-hand side. For example to store \eqref{eq:lvl2 block1 diff}, we store $\frac{\partial Q_1^{(2)}}{\partial \theta_j}$, $(B_1^{(2)T}Q_1^{(2)})$ as the two low-rank factors for the first term and $Q_1^{(2)}$,  $\left(\frac{\partial B_1^{(2)T} Q_1^{(2)}}{\partial \theta_j}\right)$ as the two low-rank factors for the second term. All these matrices are of rank $k$ resulting in a total rank of $2k$.  This computation requires $k$ K-vector products with the derivatives of the $Q$ components of the QR factorization and  $2k$ $K$-vector products with the partial derivative $K_j$, and a storage of 4 $\frac{n}{4} \times k $ blocks, or equivalently one $n \times k$ block. 

\textit{Processing the leaf level}. For the leaf level, the derivative of all the diagonal blocks can be estimated simultaneously by differentiating \eqref{eq:leaf sample},
\begin{align}
    \left(K_j - \begin{bmatrix}
    \begin{bmatrix}
    0 & \frac{\partial \hat{B}_1^{(2)}}{\partial \theta_j}\\
    \frac{\partial \hat{B}_1^{(2)T}}{\partial \theta_j} & 0
    \end{bmatrix} & \frac{\partial \hat{B}_1^{(1)}}{\partial \theta_j}\\
    \frac{\partial \hat{B}_1^{(1)T}}{\partial \theta_j} & 
    \begin{bmatrix}
    0 & \frac{\partial \hat{B}_2^{(2)}}{\partial \theta_j}\\
    \frac{\partial \hat{B}_2^{(2)T}}{\partial \theta_j} & 0
    \end{bmatrix}
    \end{bmatrix}\right)S \approx
    \begin{bmatrix}
        \frac{\partial A_1^{(2)}}{\partial \theta_j}\\
        \frac{\partial A_2^{(2)}}{\partial \theta_j}\\
        \frac{\partial A_3^{(2)}}{\partial \theta_j}\\
        \frac{\partial A_4^{(2)}}{\partial \theta_j}
    \end{bmatrix}.
    \label{eq:diag diff}
\end{align}

We access matrix $K_j$ via matrix-vector products only. We can write

\begin{align}
    K_j S = \frac{\partial (KS)}{\partial \theta_j}.
\end{align}

Recall that $S=\left[ I_{\frac{n}{4}}, I_{\frac{n}{4}}, I_{\frac{n}{4}}, I_{\frac{n}{4}} \right]^T$. We can explicitly write the second term as

\begin{align}
    \begin{bmatrix}
    \begin{bmatrix}
    0 & \frac{\partial \hat{B}_1^{(2)}}{\partial \theta_j}\\
    \frac{\partial \hat{B}_1^{(2)T}}{\partial \theta_j} & 0
    \end{bmatrix} & \frac{\partial \hat{B}_1^{(1)}}{\partial \theta_j}\\
    \frac{\partial \hat{B}_1^{(1)T}}{\partial \theta_j} & 
    \begin{bmatrix}
    0 & \frac{\partial \hat{B}_2^{(2)}}{\partial \theta_j}\\
    \frac{\partial \hat{B}_2^{(2)T}}{\partial \theta_j} & 0
    \end{bmatrix}
    \end{bmatrix}S = 
    \begin{bmatrix}
    \begin{bmatrix}
    \frac{\partial \hat{B}_1^{(2)}}{\partial \theta_j}I_{\frac{n}{4}}\\
    \frac{\partial \hat{B}_1^{(2)T}}{\partial \theta_j}I_{\frac{n}{4}}
    \end{bmatrix} + \frac{\partial \hat{B}_1^{(1)}}{\partial \theta_j} \begin{bmatrix}I_{\frac{n}{4}}\\I_{\frac{n}{4}}\end{bmatrix}\\
    \frac{\partial \hat{B}_1^{(1)T}}{\partial \theta_j} \begin{bmatrix}I_{\frac{n}{4}}\\I_{\frac{n}{4}}\end{bmatrix} + 
    \begin{bmatrix}
    \frac{\partial \hat{B}_2^{(2)}}{\partial \theta_j}I_{\frac{n}{4}}\\
    \frac{\partial \hat{B}_2^{(2)T}}{\partial \theta_j}I_{\frac{n}{4}}
    \end{bmatrix}
    \end{bmatrix}.
\end{align}
The $n \times k$ right hand side is computed using the low-rank representations of the off-diagonal blocks from \eqref{eq:diff low-rank}, \eqref{eq:lvl2 block1 diff}, and \eqref{eq:lvl2 block2 diff}, respectively. Taking the second term of the upper block for an example, we get 

\begin{align}
    \frac{\partial \hat{B}_1^{(1)}}{\partial \theta_j} \begin{bmatrix}I_{\frac{n}{4}}\\I_{\frac{n}{4}}\end{bmatrix} = \frac{\partial Q_1^{(1)}}{\partial \theta_j} \left(B_1^{(1)T}Q_1^{(1)}\right)^T \begin{bmatrix}I_{\frac{n}{4}}\\I_{\frac{n}{4}}\end{bmatrix} + Q_1^{(1)} \left(\frac{\partial B_1^{(1)T} Q_1^{(1)}}{\partial \theta_j}\right)^T \begin{bmatrix}I_{\frac{n}{4}}\\I_{\frac{n}{4}}\end{bmatrix}.
    \label{eq:tmp1}
\end{align}
The computation is done using the representation on the right by computing the rightmost factor in each term using $k^2$ inner products, then multiplying with the resulting $k \times k$ matrix the first factor in each of the two terms above. The effort is linear with the number of rows, though it is increasing linearily with level order, i.e. $O(nk^2l)$.

Therefore for each $\theta_j$, we can construct the HODLR approximation of the derivative matrix $K_j$ with off-diagonal local rank of at most $2k$ at each level. Since the complexity of differentiating the $K$-vector product using AD is linear in the complexity of evaluating the $K$-vector product itself and we need $2k$ differentiations of $K$-vector products plus $k$ extra $K$-vector products with the derivatives of the orthogonal columns at each level, the additional access to the matrix $K$ is $O(k\tau)$ $K$-vector products. In terms of extra computations, the complexity is dominated by removing all lower level off-diagonal approximations in $K$-vector products. Following the similar complexity analysis in \S \ref{sec:hodlr approx of cov matrix}, the computational complexity is given by $O(nk^2\tau^2)$. The extra storage complexity is $O(nk\tau)$ due to the off-diagonal low-rank components.

Note that we need to repeat the process for all parameters $\theta_1,\ldots,\theta_p$. In summary, to obtain the HODLR approximations of the derivative matrices of $K$ with respect to all parameters, the extra complexity is $O(pk\tau)$ $K$-vector products, $O(pnk^2\tau^2)$ time and $O(pnk\tau)$ memory.

Similarly if assuming constant off-diagonal rank $k=O(1)$ and the number of levels grows as $O(\log{n})$, for example $\tau=\lfloor\log_2{(\frac{n}{k})}\rfloor$, the computational complexity is $O(p\log{n})$ $K$-vector products and $O(pn\log^2{n})$ complexity. The storage complexity is $O(pn\log{n})$.

\section{Hierarchical Approximations of Gaussian Likelihood, Score Equations, Information Matrices}
\label{sec:HODLR approx of likelihood, scores, info matrix}

With the HODLR approximations of the covariance matrix and its derivatives at hand, we can then efficiently approximate the Gaussian log-likelihood function, the score equations and the observed Fisher information matrix.

\subsection{Approximated Gaussian Likelihood}

Assuming a constant rank $k$, the HODLR approximation $\tilde{K}$ requires $O(\log{n})$ levels. Such a structure admits an exact factorization with $O(n\log^{2}{n})$ computational complexity \cite{sym-factorization-of-hodlr}. The resulting approximation of the exact log-likelihood function defined in \eqref{eq:log-like} is denoted by $\tilde{L}(\theta)$:
\begin{align}
    \tilde{L}(\theta) & =  -\frac{n}{2}\log(2\pi)-\frac{1}{2}\log{|\tilde{K}|}-\frac{1}{2}(y-\bar{y})^T \tilde{K}^{-1}(y-\bar{y}).
    \label{eq:log-like approximated}
\end{align}
As we discussed in Section \ref{sec:HODLR Matrices}, the determinant and linear system can both be solved in $O(n\log^2{n})$ complexity. Therefore given the factorization of $\tilde{K}$, the approximate log-likelihood \eqref{eq:log-like approximated} can be evaluated in $O(n\log^2{n})$ time. 

\subsection{Trace Computation}
\label{new operations for hodlr}

When evaluating the score equations, since both $K$ and $K_j$ have been approximated in HODLR format, the second term in \eqref{eq:score} poses no difficulty. The bottleneck is obviously the trace of matrix products. 
One option for it is the use of stochastic trace estimators which converts the trace operation to a sequence of matrix-vector products with random vectors. Several choices are available including Gaussian trace estimator, Hutchinson's trace estimator, unit vector trace estimator; we refer readers to \cite{trace-estimator} for a discussion and convergence analysis. Given a relative error tolerance $\epsilon$, generally it requires $O(\epsilon^{-2})$ random samples to achieve that   accuracy.

Another option is to solve the matrix product explicitly. As we discussed in Table \ref{sec:HODLR Matrices}, computing $A^{-1}B$ for two HODLR matrices $A,B$ have been studied and can be conducted in $O(n\log^2{n})$ time. However as arithmetic operations often increase the HODLR ranks, it is a common practice to combine the operations with recompression and perform low-rank truncation for each off-diagonal block to control the local rank growth. However in this case the operations incur additional error (which may be acceptable given the HODLR approximation error itself but we would like to explore avoiding it).

The third option is to take the advantage of the properties of the trace operation to simplify the computations. Based on the expression of \eqref{eq:score} and \eqref{eq:fisher info} we introduce two operations $\mathrm{tr}(A^{-1}B)$ and $\mathrm{tr}(A^{-1}BC^{-1}D)$ given $A,B,C,D$ HODLR matrices. Both operations can be conducted exactly and in quasilinear scale. The detailed algorithms are included in Appendix \ref{app:two new operations}. In this work, we will use the third option when computing the score equations and Fisher information matrix associated with the approximate log-likelihood   \eqref{eq:log-like approximated}.

\subsection{Approximated Score Equations And Fisher Information Matrix}

Now consider the score equations \eqref{eq:score}. The score equations of the approximate likelihood are given by
\begin{align}
    \tilde{S}_j(\theta)&=-\frac{1}{2}\mathrm{tr}\left(\tilde{K}^{-1}\tilde{K}_j\right) + \frac{1}{2}y^T \tilde{K}^{-1} \tilde{K}_j \tilde{K}^{-1}y.
    \label{eq:score approximated}
\end{align}
Recall $\tilde{K}_j$ is also an HODLR matrix with rank $3k$  \eqref{eq:lvl2 block1 diff}. Utilizing the proposed trace operation in \ref{app:two new operations}, we can now evaluate the trace term exactly in $O(n\log^2{n})$ scale. For the second term we compute $\tilde{K}^{-1}y$ using the factorization of $\tilde{K}$ and the rest are HODLR matrix-vector products. Repeat the computation for all $p$ parameters and we get that the approximated score equations \eqref{eq:score approximated} can be evaluated in an extra $O(pn\log^2{n})$ time.

With the ability to efficiently approximate the log-likelihood and the score equations, we can then apply optimization algorithms to obtain the maximum likelihood estimates of the parameters. Note that the parameter values are updated at each optimization iteration. Therefore new HODLR approximations need to be constructed for every iteration. Once the MLE estimator $\hat{\theta}$ for the parameters is obtained, the next step is to approximate the observed Fisher information matrix for uncertainty quantification. Denote the approximated Fisher information matrix by $\tilde{\mathcal{I}}$. Its entries are given by
\begin{align}
    \tilde{\mathcal{I}}_{i,j}(\hat{\theta}) &= \frac{1}{2}\mathrm{tr}
    \left[{\left(\tilde{K}^{-1} \tilde{K}_i \tilde{K}^{-1} \tilde{K}_j\right)} |_{\theta=\hat{\theta}}\right].
    \label{eq:fisher info approximated}
\end{align}

Note here all the matrices are evaluated at the MLE estimator $\hat{\theta}$. Based on the proposed operations in \ref{app:two new operations}, given the factorization of $\tilde{K}$ and HODLR matrices $\{\tilde{K}_j\}_{j=1,\cdots,p}$, evaluating each entry of the Fisher information matrix takes $O(n\log^2{n})$ time. Repeating for all entries, a total $O(p^2n\log^2{n})$ complexity is required.

The number of parameters to estimate is often smaller comparing to the size of the observations. In summary, each optimization iteration takes $O(p\log{n})$ $K$-vector products to construct the HODLR approximations and $O(pn\log^2{n})$ operations to carry out. After the optimization process, estimating the Fisher information matrix takes $O(p^2n\log^2{n})$ time. If the $K$-vector product number of operations is quasi-linear in $n$ as well (which, for implicitly defined covariance models with sparse state space operators e.g \eqref{eq:latent covariance}, is the case in many circumstances), then our approach has \textit{quasillinear effort in building the covariance matrices and its derivatives, to factorize it, and per iteration of max-likelihood computation.} 

\section{Numerical Experiments}
\label{sec:numerical experiments}

We perform several numerical experiments with synthetic datasets to demonstrate the scaling of our approach and the ability to recover the true values of the parameters from the data. We provide numerical evidence for both the computational scaling and accuracy of the parameter estimations.
We note that our approach in \S \ref{sec:cov model and MLE}
allows deriving covariance functions from physical principles followed by statistical analysis that employs the entire data set, the latter underpinning our objective of quasilinear performance. It is worth asking whether such a computational effort is worth the significant development cost and whether one can get away with fitting with only a portion of the data. To this end, we present in Appendix \S \ref{sec:motivating} an example of fitting a nonstationary process with a spatially linearily changing lengthscale. In that example we observe that fitting only on a subdomain results in good confidence intervals for the intercept of the lengthscale model but not the slope, whereas subsampling the data reverses that behavior, and, finally, using the entire data set produces good confidence intervals for both parameters, Figure \ref{fig:toy_mle}. Our approach in \S \ref{sec:cov model and MLE} allows for much more flexible modeling, including complex nonstationary models by, for example, using PDEs with spatially dependent parameters. It is thus likely to result in models where the dependency of those parameters on the lengthscales is complex and deciding \textit{which} portion of the data to use for the purpose of obtaining good estimates may be a difficult endeavor. To this end, we conclude that deriving algorithms which allow the use of the entire data set may be worthwhile, and it is in this vein that we present the following numerical experiments. 

\subsection{SPDE Representation of Mat\'ern Models}
\label{sec:SPDE approach for Matern}

The SPDE approach to Gaussian fields can significantly reduce the computational cost of inference and sampling
by invoking a GMRF approximation. One of the most popular examples is the stationary Mat\'ern model. Recall a Gaussian field belongs to the Mat\'ern family if its covariance function can be written in the form
\begin{align}
    M_{\nu,l}\left(\mathbf{x},\mathbf{y}\right) = (2^{\nu-1}\Gamma(\nu))^{-1}\left(\frac{||\mathbf{x}-\mathbf{y}||_2}{l}\right)^\nu K_\nu\left(\frac{||\mathbf{x}-\mathbf{y}||_2}{l}\right).
    \label{eq:whittle covariance}
\end{align}
where $\Gamma$ denotes the Gamma function and $K_\nu$ denotes the
modified Bessel function of the second kind. The smoothness parameter $\nu$ controls the regularity of the random field, i.e. the degree of differentiability. An important characterization by Whittle \cite{whittle1,whittle2} is that the Mat\'ern fields can be be defined as the solution to certain fractional order stochastic partial differential equation. Specifically, a Gaussian field with covariance of form $\sigma_m^2 M_{\nu,l}\left(\mathbf{x},\mathbf{y}\right)$ is the unique stationary solution to the SPDE 
\begin{align}
    \left(\frac{1}{l^2} - \Delta\right)^{\frac{\nu+d/2}{2}} (\gamma w(\mathbf{x})) = \mathcal{W}(\mathbf{x}),\ \mathbf{x}\in\mathbb{R}^d,
    \label{eq:spde of matern}
\end{align}
where $\mathcal{W}$ denotes the spatial Gaussian white noise with unit variance and the marginal variance of $w$ is given by
\begin{align}
    \sigma_m^2 = \frac{\Gamma(\nu)l^{2\nu}}{\Gamma(\nu+d/2)(4\pi)^{d/2}\gamma^2}. 
\end{align}
Therefore we can control $\gamma$ to get desired marginal variance for the Mat\'ern field. It is worth pointing out that such models have sparse inverse covariance matrices, and, for example, the log-likelihood can be easily computed in quasilinear time directly without needing our approach \cite{link-spde-matern} . The score equations, however, and in particular, the trace term of \eqref{eq:score} do not have an obvious way to compute in quasilinear time if one wants to pursue maximum likelihood calculations. It is worth then investigating the potential benefit of using our approach to this end. 

As discussed in \cite{link-spde-matern}, the SPDE formulation of Mat\'ern field allows to simulate the random field by computing the solution of \eqref{eq:spde of matern}. And \eqref{eq:spde of matern} can further be efficiently approximated by finite element analysis. In all the numerical simulations and studies described below, we fix the smoothness parameter $\nu=1$ and restrict our attention to 2D space $d=2$. In this case the order of the SPDE differential operator is an integer $\frac{\nu+d/2}{2}=1$. Now the SPDE becomes
\begin{align}\label{eq:SPDE}
    \left(\frac{1}{l^2} - \Delta\right) (\gamma w(\mathbf{x})) = \mathcal{W}(\mathbf{x}),\ \mathbf{x}\in\mathbb{R}^d.
\end{align}

Given a set of finite element basis functions $\{\Phi_i(\mathbf{x})\}_{i=1,\cdots,N_b}$, we solve \eqref{eq:SPDE} via standard finite element analysis. Assume that we observe the random process $w$ at $n$ observation points $(\mathbf{x}_1,\cdots,\mathbf{x}_n)$. We construct the observation matrix by evaluating the basis functions at given observation locations,

\begin{align}
    \mathbf{\Phi} = 
    \begin{bmatrix}
        \Phi_1(\mathbf{x}_1) & \Phi_2(\mathbf{x}_1) & \cdots & \Phi_{N_b}(\mathbf{x}_1)\\
        \vdots & \vdots & \ddots & \vdots\\
        \Phi_1(\mathbf{x}_n) & \Phi_2(\mathbf{x}_n) & \cdots & \Phi_{N_b}(\mathbf{x}_n)
    \end{bmatrix}.
\label{eq:projection}
\end{align}
Note that $\Phi$ needs not be square and this flexibility can be used to define a latent grid on which the operators can be easily inverted (e.g a rectangle), but use the projection matrix to map onto the data space. Nevertheless we will assume that matrix vector products with $\Phi$ and $\Phi^T$ are easy to carry out, e.g the observations are well spread out and $\Phi$ is sparse.

Denote the finite element mass matrix $C_{ij}=\int_{\mathbb{R}^2}\Phi_i(\mathbf{x})\Phi_j(\mathbf{x}) d\mathbf{x}$ and stiffness matrix $S_{ij}=\int_{\mathbb{R}^2} \nabla\Phi_i(\mathbf{x}) \cdot \nabla\Phi_j(\mathbf{x}) d\mathbf{x}$. Further let $\tilde{C}$ denote the extracted diagonal matrix from $C$. The approximated finite-dimensional random field follows multivariate Gaussian distribution,
\begin{align}
    (w(\mathbf{x}_1),\cdots,w(\mathbf{x}_n))^T \sim \mathcal{N} \left(0, \mathbf{\Phi} K_w \mathbf{\Phi}^T \right),\ \text{where }K_w=\frac{1}{\gamma^2}  \left(\frac{1}{l^2}C+S\right)^{-1} \tilde{C} \left(\frac{1}{l^2}C+S\right)^{-T}.
    \label{eq:latent covariance}
\end{align}

More details about the finite element analysis can be found in \cite{rational-spde-approx}. Note that the covariance structure of \eqref{eq:latent covariance} consists of a sequence of products and matrix inverses of sparse finite element matrices. This model form enables fast covariance matrix-vector products through sequential sparse matrix-vector products.

\subsection{A Mat\'ern Based Wind Velocity Model}\label{sec:wind}

Here we use a Mat\'ern based Gaussian process model for the horizontal wind components proposed in \cite{wind-model} as our numerical model. Specifically the horizontal wind velocity has two components $U=(u,v)^T$. The two components are connected via the Helmholtz decomposition, which states that for any given wind field $U$ there exists a streamfunction $\phi$ and velocity potential $\chi$, such that $U=\nabla\times\phi + \nabla\chi$. Assume the streamfunction and the velocity potential have the following bivariate Mat\'ern structure:

\begin{align}
    K_{\phi,\chi}(\mathbf{x},\mathbf{y}) = 
    \begin{bmatrix}
        \sigma^2_{\phi} & \rho\sigma_\phi\sigma_\chi\\
        \rho\sigma_\phi\sigma_\chi & \sigma^2_{\chi}
    \end{bmatrix}
    M_{\nu,l}\left(\mathbf{x},\mathbf{y}\right),
    \label{eq:analytical latent covariance}
\end{align}
where $\mathbf{x},\mathbf{y}$ are 2D spatial locations of the random field and $M_{\nu,l}$ denotes the Whittle covariance function given by \eqref{eq:whittle covariance}. 

To approximate the bivariate Mat\'ern covariance matrix \eqref{eq:analytical latent covariance} using the SPDE approach \S \ref{sec:SPDE approach for Matern}, we assume that both the streamfunction and the velocity potential are observed at the same set of locations. Then we have
\begin{align}
    K_{\phi,\chi} \approx \begin{bmatrix}
        \sigma^2_{\phi} & \rho\sigma_\phi\sigma_\chi \\
        \rho\sigma_\phi\sigma_\chi  & \sigma^2_{\chi}
    \end{bmatrix} \otimes K_w,
\label{eq:kphichi}
\end{align}
where $\otimes$ denotes the Kronecker product and $K_w$ comes from \eqref{eq:latent covariance}. 

The physics-based covariance model discussed back in \secref{sec:physics-based cov model} provides a flexible way to construct the covariance model of the wind velocity components using underlying physics relations. Here we restate the physics model involving the three random fields,
\begin{align}
    U = (u,v)^T = \nabla\times\phi + \nabla\chi = \left(-\frac{\partial}{\partial e_2}\phi + \frac{\partial}{\partial e_1}\chi, \frac{\partial}{\partial e_1}\phi + \frac{\partial}{\partial e_2}\chi\right)^T.
    \label{eq:wind model}
\end{align}

We can use finite difference method to discretize the operator or we can directly take the derivative of $(\phi,\chi)$ since we have numerically approximated the solutions using finite element basis functions. To enable fast matrix-vector products, we take the finite difference discretization of the differential operators. Let $L_{1,2}$ denote the discretized one-dimensional differential operator with respect to the two directions $e_1$ and $e_2$. Then we have the discretized model
\begin{align}\label{eq:}
    U = 
    \begin{bmatrix}
        -L_2 & L_1\\
        L_1 & L_2
    \end{bmatrix}
    \begin{bmatrix}
        \phi\\ \chi
    \end{bmatrix} = L\begin{bmatrix}
        \phi\\ \chi
    \end{bmatrix}.
\end{align}
Now we use the physics-based covariance model in \eqref{eq:cov model}, the covariance of the wind velocity components can be written as
\begin{align}
    K_U =\sigma_n I_{2n}+ \hat{\Phi}LK_{\phi,\chi}L^T\hat{\Phi^T}.
    \label{eq:output cov model}
\end{align}
Here $\hat{\Phi}$ is a latent-to-data projection operator, and we assume that the measurements are noisy with known variance $\sigma_n$ (selected so the variance is a fraction of the sample variance). If $u,v$ are available at the same points, then we can have $\hat{\Phi}=[ \Phi^T, \Phi^T]^T$, where $\Phi$ is the interpolation operator from 
\eqref{eq:projection} (which is what we will use in our calculations). Since $L$ is obtained by divided differences, it is sparse, moreover, we have access to efficient solvers for applying the inverse operators in $K_w$ \eqref{eq:latent covariance} needed to access matrix-vector products with $K_{\phi,\chi}$  \eqref{eq:kphichi}. 

In building $K_U$ we have shown one way to construct a meaningful, moderately complex model which has the feature that matrix-vector products with $K_U$ can be evaluated fast $O(n \log(n))$ in this case (assuming the latent space size is of the order of the data space size $n$), \textit{despite the fact that $K_U$ itself is dense. Therefore, storing it may be inaccessible for large data sets and thus even more so having access to its Cholesky factors needed for the log-determinant term  in the likelihood.}  A similar construction pattern can be applied beyond the classical Mat\'ern model to circumstances where the SPDE representation can also be extended to non-stationary, non-isotropic fields \cite{link-spde-matern}, temporal processes \cite{temporal-spde} and spatial-temporal processes \cite{spatial-temporal-spde}, and the resulting model will have the same property. 

For our experiments, we use \eqref{eq:output cov model} and fix the smoothness parameter to $\nu=1$ in the  Mat\'ern covariance function which corresponds to a process that is just barely not mean-square differentiable. As discussed in \cite{wind-model}, realistic mesoscale wind fields have a smoothness parameter close to $1.25$ to which our model is close. We use as true parameter values $(\rho,\sigma_\phi,\sigma_\chi,l)=(0.7,1,0.3,0.5)=\theta_{\mathrm{true}}$ and simulate five datasets on the 2D domain $[-5,5]^2$ from the Mat\'ern field using \textit{R} package \textit{RandomFields} of \cite{R-random-field}. Each dataset contains $2^{20}$ observations for both $\phi$ and $\chi$ on an even grid across the domain; the wind field is then generated by the relation \eqref{eq:wind model}. We also note that to obtain the approximation of the Mat\'ern covariance via \eqref{eq:latent covariance}, we perform finite element analysis with Neumann boundary condition on a slightly larger domain $[-5.5,5.5]^2$ to avoid the boundary effect. 

To approximate the covariance matrix by an HODLR format we use leaf level blocks with sizes $256$ or $512$. Therefore, the maximum number of levels of the HODLR approximation is $\left \lfloor{\log_2{(n/256)}}\right \rfloor $ where $n$ is both the size of $K_U$ and number of observations of the wind field $U$.  For all off-diagonal blocks, a fixed rank is used which we will experiment with and specify later. The ordering of the observations is done by a KD-tree binary spatial partitioning. The reordering step increases the degree to which
off-diagonal blocks correspond to the covariance between groups of well-separated points, a necessary characteristic for an efficient hierarchically low rank representation. 
All the following computations shown in this section were performed on a standard workstation with a 2.4 GHz CPU and 32 GB of memory.

\subsubsection{Quasilinear Scaling of the Log-likelihood, Score Equations and the Fisher Information}

Our datasets are generated over a $2^{10}\times 2^{10}=2^{20}$ grid at the finest level. To demonstrate the numerical scaling, we coarsen the grid while also coarsening the finite element grid in \eqref{eq:latent covariance} and the PDE operator $L$ by the same factor. We then subsample observations according to the coarsened grids from the original dataset. For example, by coarsening both directions using a factor of $2$, we can get a $2^{9}\times 2^{9}=2^{18}$ grid. Then we subsample the generated observations of $\phi,\chi,U$ according to the coarse grid. Now we obtain a dataset containing $2^{18}$ data points. Performing the same downsampling procedure again provides us a dataset with $2^{16}$ data points. Additionally we can randomly sample half of the grid points to observe on the $2^{18}$ grid. This gives us $2^{17}$ observations. Using the method described above, we generate datasets of size $2^r$ on irregular grids for any $r=1,\cdots,20$ by properly coarsening the grids and taking partial observations.

We generate subsampled datasets of sizes $2^r$ for $r$ ranging from $9$ to $16$. The construction of the HODLR approximation of $K_U$ is based on its products with sampling vectors as described in \ref{sec:hodlr approx of cov matrix}. Here we test three different ranks for HODLR off-diagonal blocks $k=32,64,128$.

\begin{figure}[htb!]
\centering
\begin{subfigure}[t]{0.4\textwidth}
    \includegraphics[width=\textwidth]{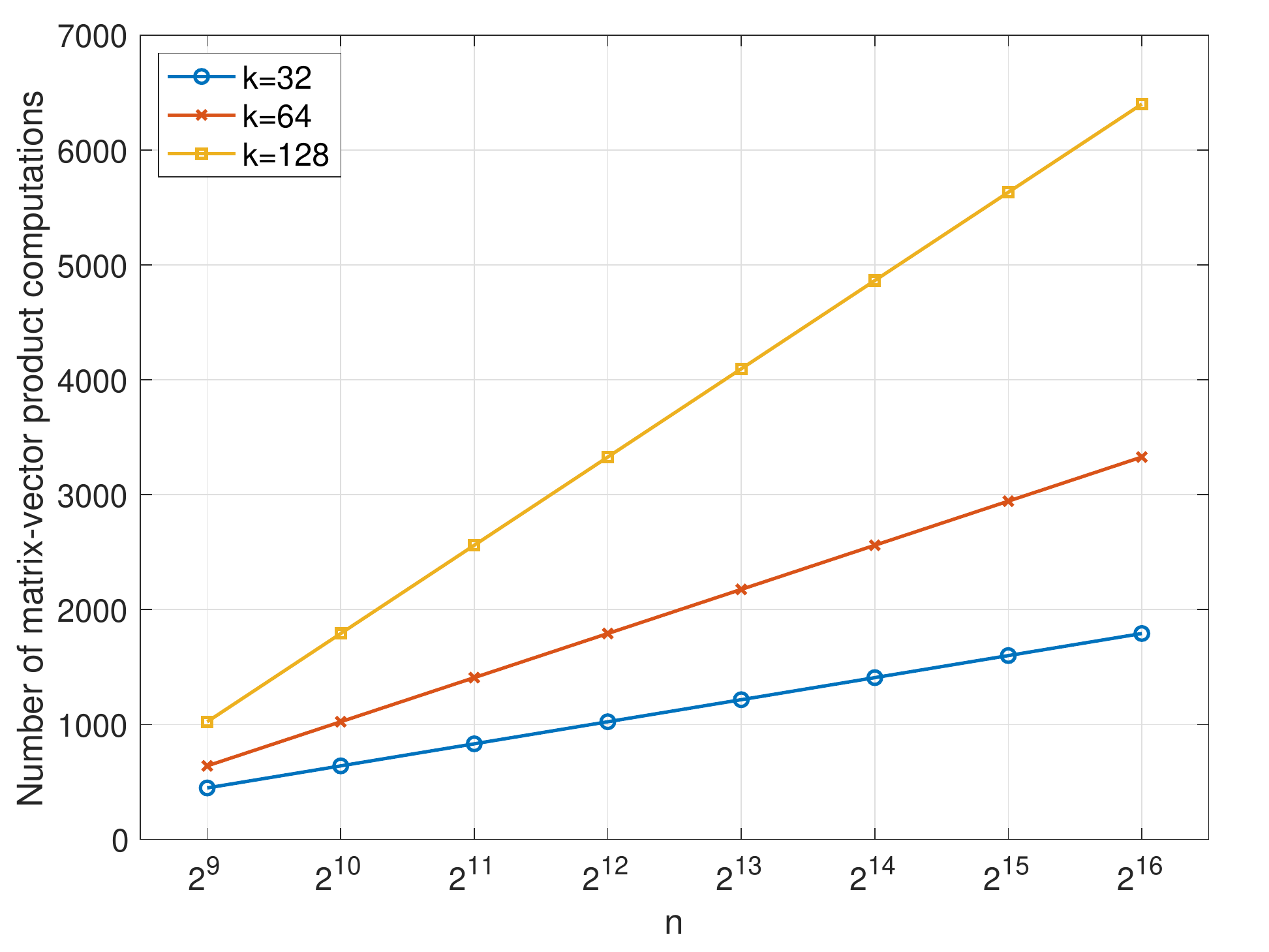}
    \caption{Number of matrix-vector product evaluations}
\end{subfigure}
\begin{subfigure}[t]{0.4\textwidth}
    \includegraphics[width=\textwidth]{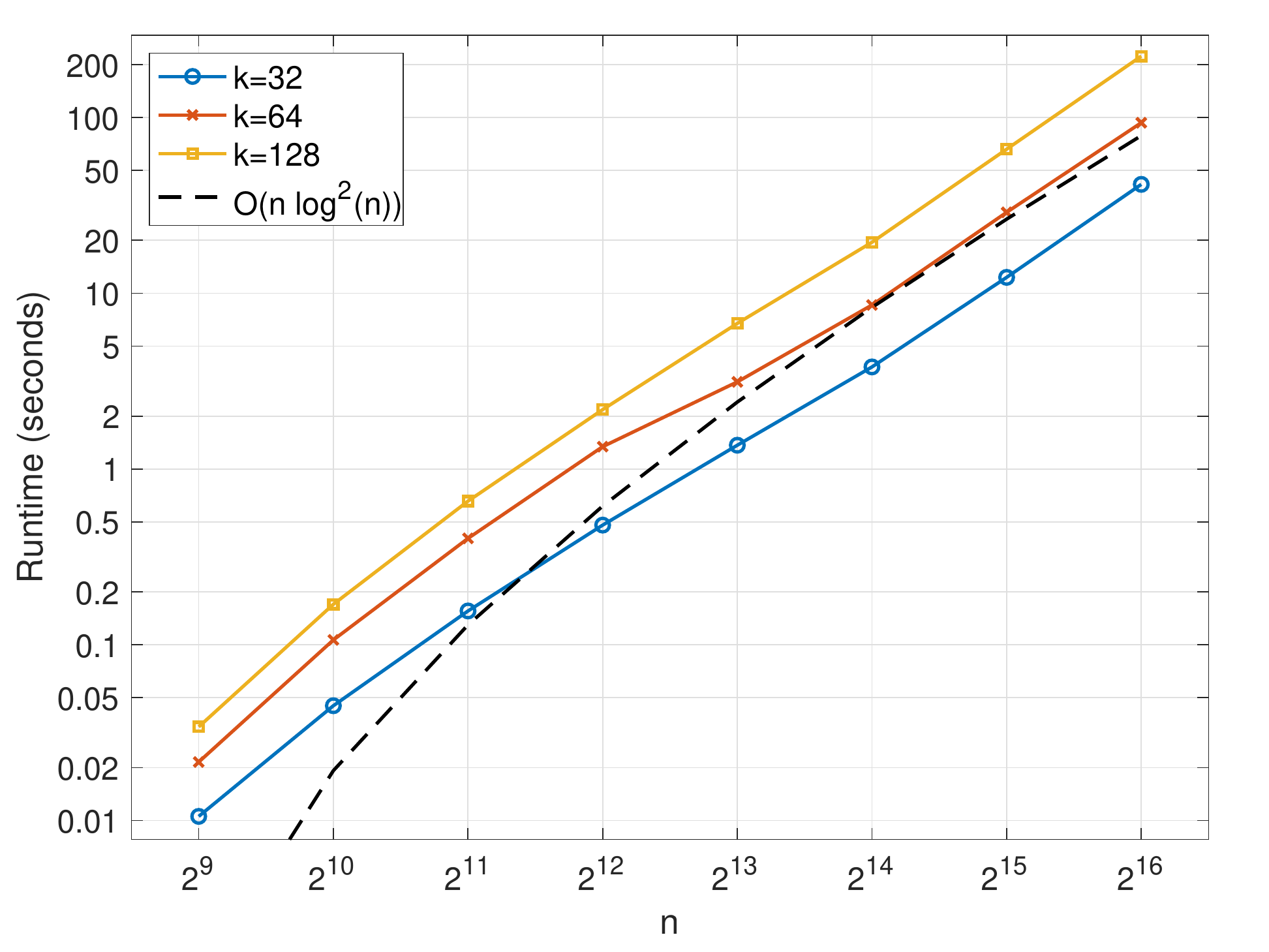}
    \caption{Runtime of rest of the operations}
\end{subfigure}
\caption{Computational complexity of constructing the HODLR approximation of $K_U$ and its derivatives with respect to the parameters $\theta$: (a) the total number of required $K_U$-vector products and (b) shows the runtime (in seconds) of the rest linear algebra operations for fixed off-diagonal rank 32 (blue curve with circles), 64 (red curve with crosses), 128 (yellow curve with squares) over different sizes of observations. We use number of observations of size $n=2^r$ with $r$ ranging from $9$ to $16$. In (b), to demonstrate the scaling, the theoretical line (black dashed line) corresponding to $O(n\log^2{n})$ is added to the plot.}
\label{fig:hodlr cost}
\end{figure}

Following our proposed workflow, we first construct the HODLR approximation for $K_U$ and all its derivatives with respect to parameters $\theta=(\rho, \sigma_\phi, \sigma_\chi, l)$ by matrix-vector products. The time complexity mainly comes from two parts: the evaluations of matrix-vector products with sampling vectors and the linear algebra computations including the QR factorization. Thus we count the number of covariance matrix-vector evaluation calls and record the time spent on the remaining linear algebra computations. The results are summarized in \figref{fig:hodlr cost}.

As can be seen in \figref{fig:hodlr cost}, the number of matrix-vector products is linear with $\log{n}$ and the slope depends on the off-diagonal rank we use. The scaling of the rest operations exactly follows the expected $O(n\log{n})$ scale.  This validates our complexity estimation. 

\begin{figure}[htb!]
\centering
\begin{subfigure}[t]{0.4\textwidth}
    \includegraphics[width=\textwidth]{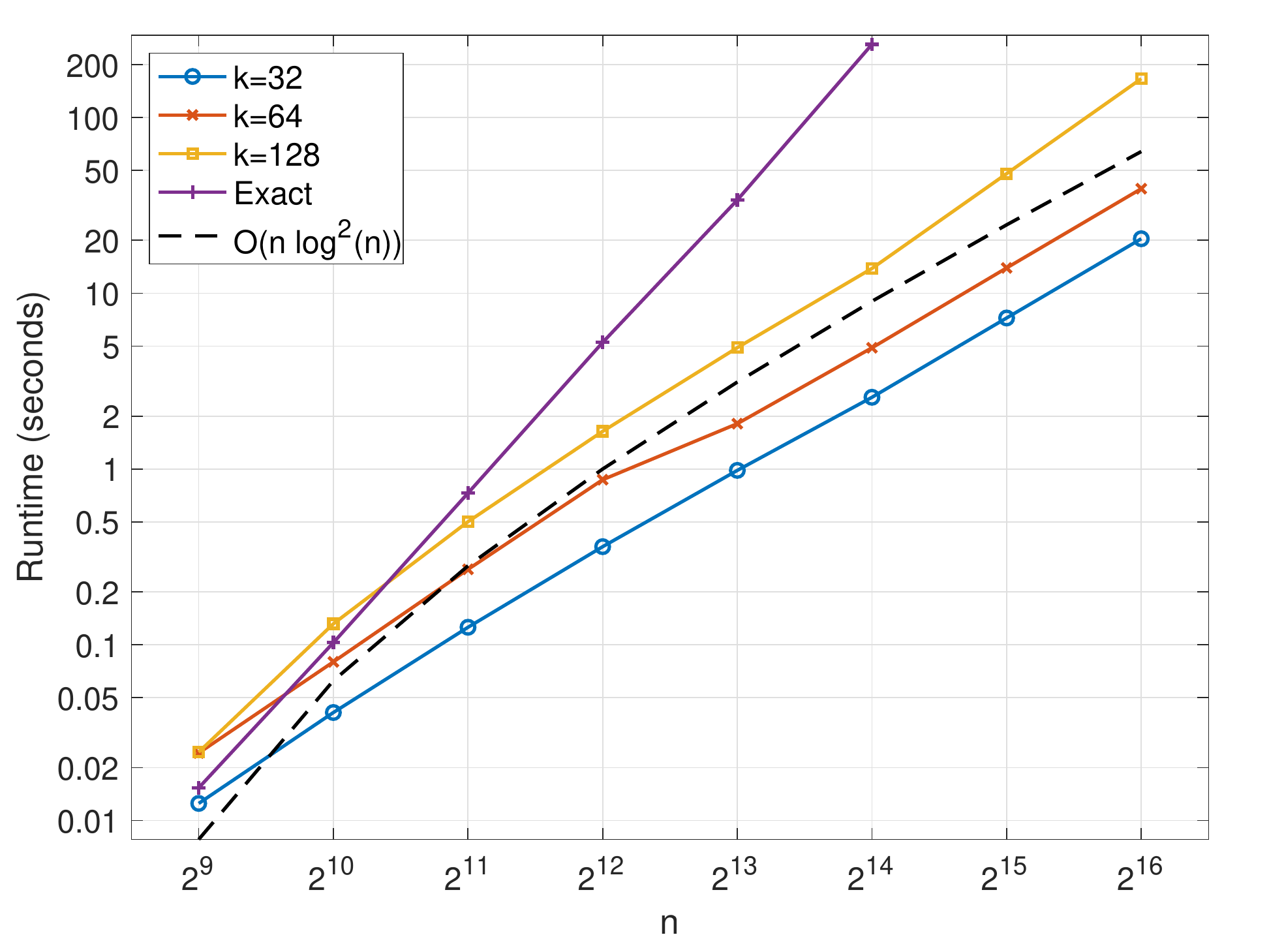}
    \caption{$\mathrm{tr}(A^{-1}B)$}
\end{subfigure}
\begin{subfigure}[t]{0.4\textwidth}
    \includegraphics[width=\textwidth]{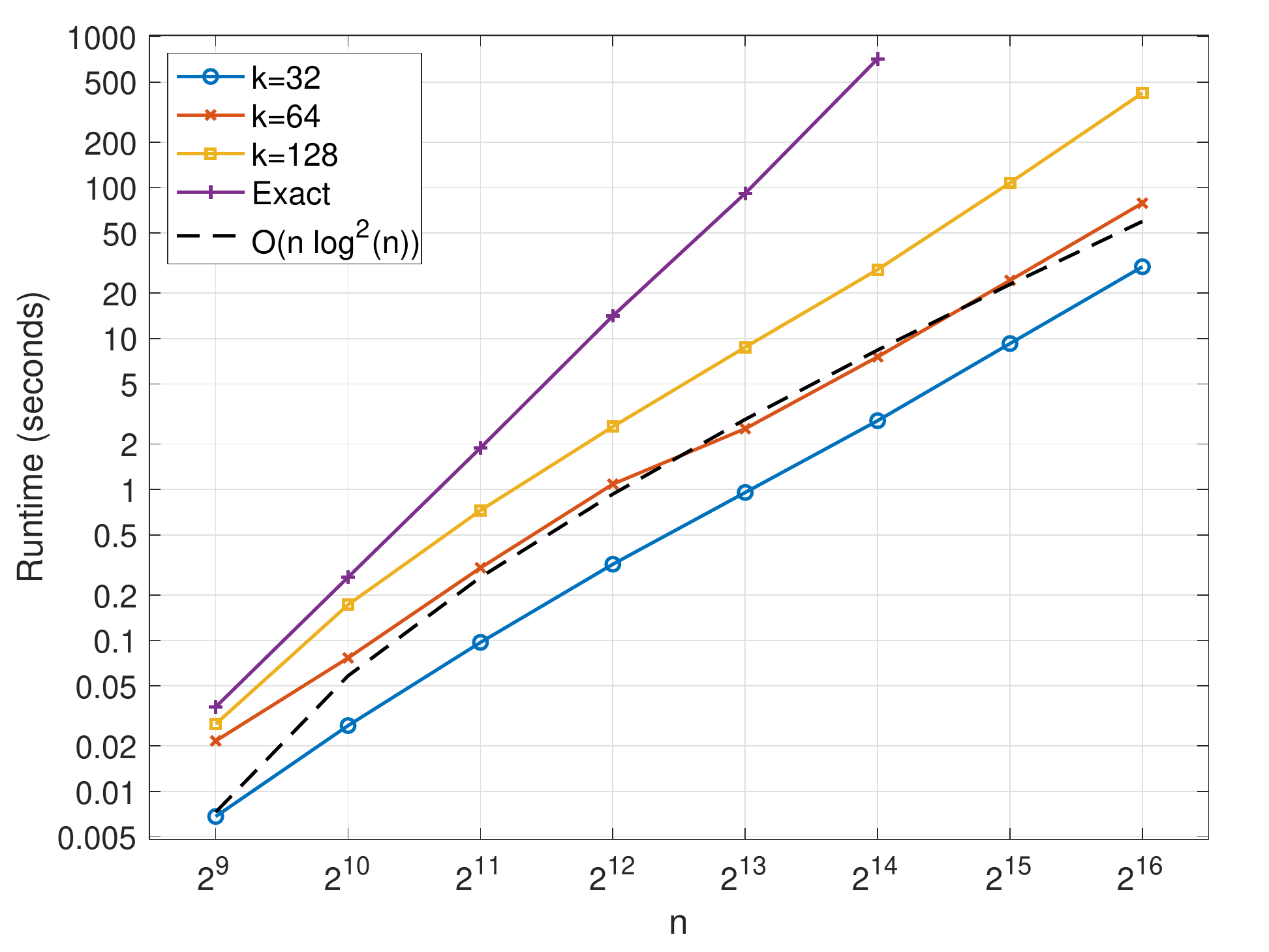}
    \caption{$\mathrm{tr}(A^{-1}BC^{-1}D)$}
\end{subfigure}
\caption{Runtime (in seconds) of (a) $\mathrm{tr}(A^{-1}B)$ given two HODLR matrices $A,B$ and $\mathrm{tr}(A^{-1}BC^{-1}D)$ given the two products $A^{-1}B$ and $C^{-1}D$ for fixed off-diagonal rank 32 (blue curve with circles), 64 (red curve with crosses), 128 (yellow curve with squares) over $n=2^r$ observations with $r$ from $9$ to $16$. Theoretic lines corresponding to $O(n\log^2{n})$ scaling (purple curve with pluses) are added to each plot. Additionally, we include the complexity of their corresponding exact operations (black dashed line) when $A,B,C,D$ are dense and have no structure in each plot for $r\leq 14$. }
\label{fig:trace cost}
\end{figure}

Next, as the fundamental building blocks of approximating the score equations and Fisher information matrix, we want to show the two new operations $\mathrm{tr}(A^{-1}B)$ and $\mathrm{tr}(A^{-1}BC^{-1}D)$ proposed in Appendix \ref{app:two new operations} are indeed quasilinear scale. In \figref{fig:trace cost}, we demonstrate the complexity of both operations (a) $\mathrm{tr}(A^{-1}B)$ and (b) $\mathrm{tr}(A^{-1}BC^{-1}D)$ given HODLR matrices $A,B,C,D$.  As can be seen above, the computational scaling of both operations follows closely the expected $O(n\log^2{n})$ line. We also compare the results with their exact counterpart when the matrices are dense and have no hierarchical structures for $r\leq 14$, as can be seen our approach is  $10-20$ time faster at $r=14$ at which point the exact approach runs out of memory. 

We now evaluate the complexity of computing the approximated log-likelihood, score equations and observed Fisher information matrix. Given the constructed HODLR approximation of the covariance matrix $K_U$ and its derivatives with respect to all the parameters, we show the time taken to evaluate the log-likelihood, the score equations and the Fisher information matrix in \figref{fig:like, score, fisher}. All the three evaluations follow the theoretical $O(n\log^2{n})$ scale, which is also indicative of the complexity of the time per iteration when we use the log-likelihood and score equations in a maximum likelihood algorithm. We conclude that our HODLR approximation algorithm for  the log-likelihood, the score equations and the Fisher information matrix has quasilinear complexity and, moreover, the number of true covariance matrix-vector products required  to build the approximation is $O(\log{n})$, as we claimed in our analysis. 

\begin{figure}[htb!]
\centering
\begin{subfigure}[t]{0.31\textwidth}
    \includegraphics[width=\textwidth]{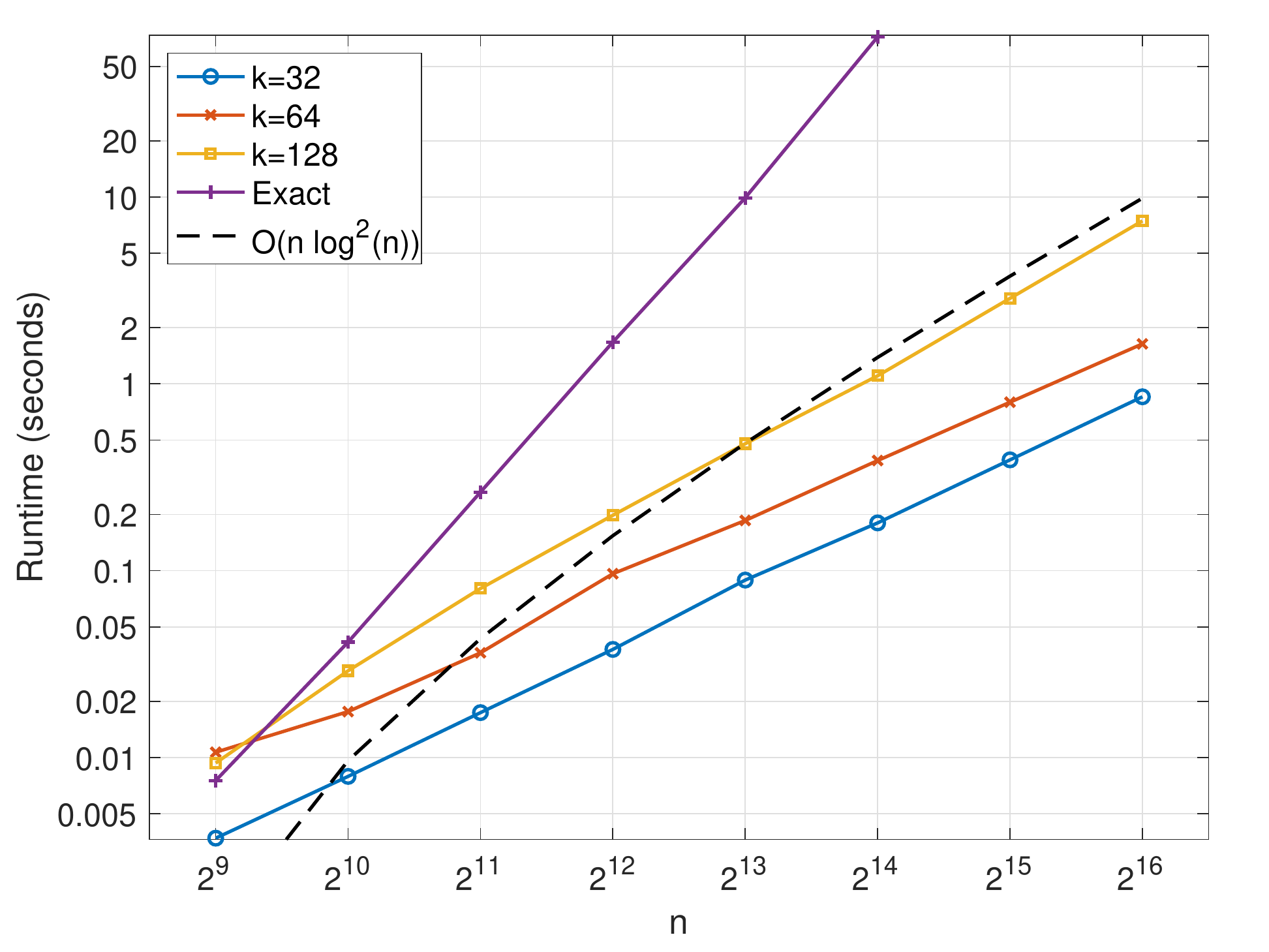}
    \caption{Log-likelihood}
\end{subfigure}
\begin{subfigure}[t]{0.31\textwidth}
    \includegraphics[width=\textwidth]{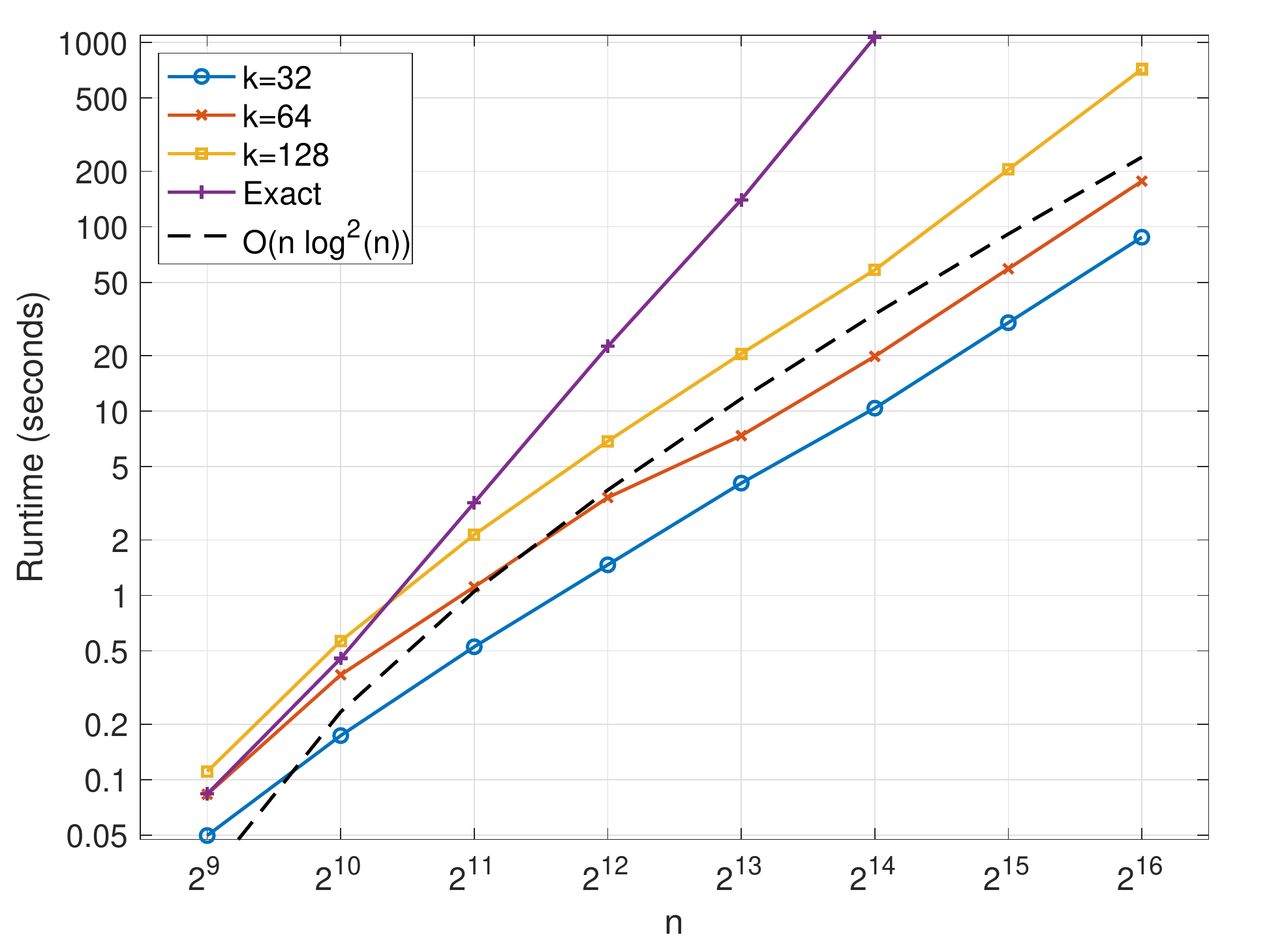}
    \caption{Score equations}
\end{subfigure}
\begin{subfigure}[t]{0.31\textwidth}
    \includegraphics[width=\textwidth]{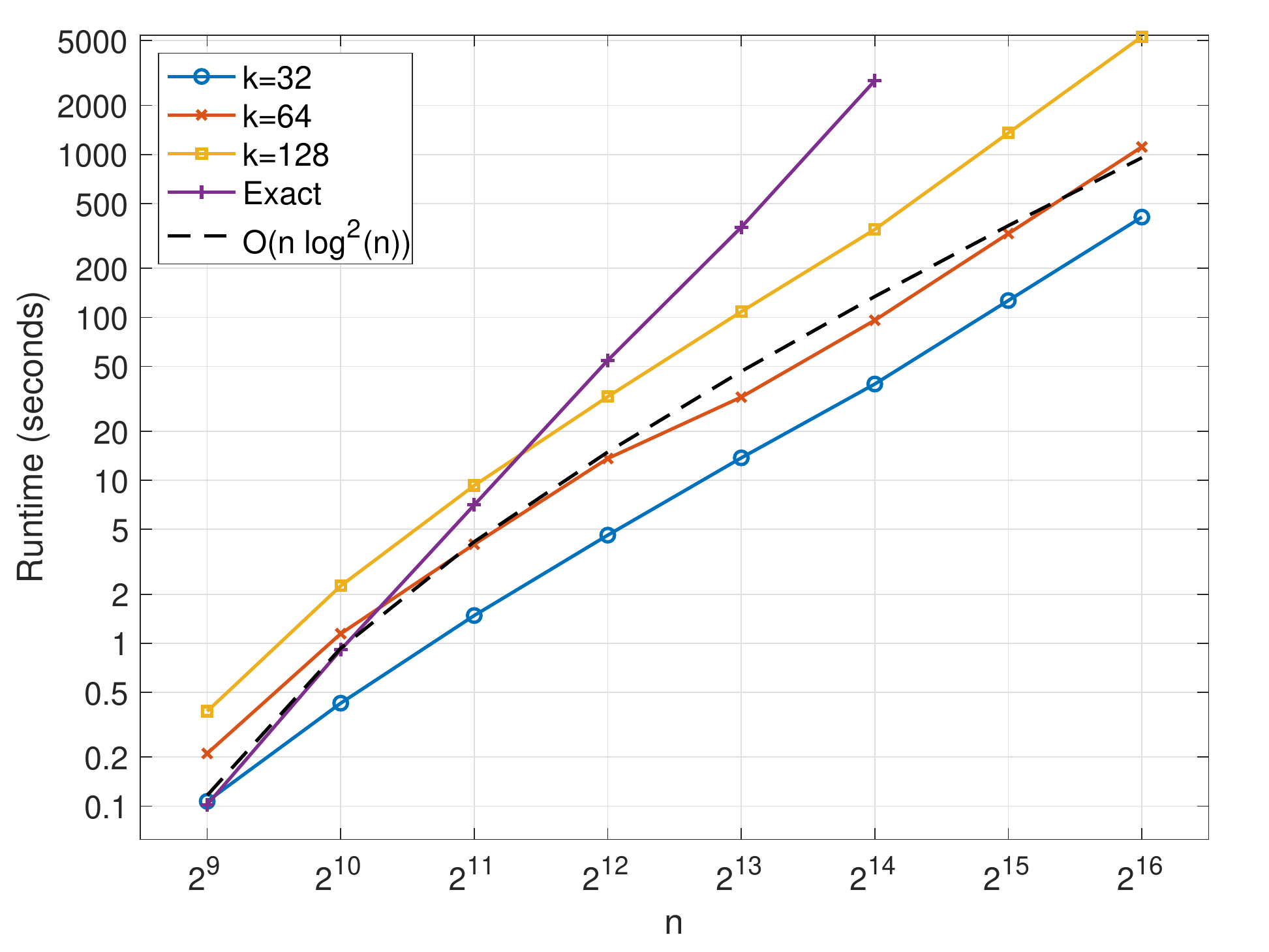}
    \caption{Observed Fisher information matrix}
\end{subfigure}
\caption{Time taken (in seconds) to evaluate (a) the log-likelihood, (b) the score equations and (c) the observed Fisher information matrix exactly (purple curve with pluses) and using HODLR approximations for fixed off-diagonal rank 32 (blue curve with circles), 64 (red curve with crosses), 128 (yellow curve with squares) over $n=2^r$ observations with $r$ from $9$ to $16$. Theoretic lines corresponding to $O(n\log^2{n})$ scaling (black dashed line) are added to each plot.}
\label{fig:like, score, fisher}
\end{figure}

\subsubsection{Numerical Accuracy of the Log-likelihood, Score Equations and the Fisher Information}

Next we demonstrate the numerical accuracy of the approximated log-likelihood, score equations and Fisher information matrix comparing to their exact values. When carrying out the computations we will compare our model with another popular low-rank-type model, \textit{the sparse spiked covariance model} \cite{sparse_spiked}, which can be thought of as diagonal plus low rank model. This will allow us to assess whether the workflow complexity and thus the development complexity of our model, which, as can be seen above, is considerable, is justified compared to the benefits it brings.

Sparse spiked covariance model for this model approximation class,  we approximate the true covariance matrix by diagonal plus low-rank components \cite{sparse_spiked}:
	\begin{align} \label{eq:spiked}
		K_U \approx \sigma_n I_{2n} + VV^T.
	\end{align}
	Here $\sigma_n$ is the same as in \eqref{eq:output cov model}, and the approach is equivalent to considering the "physics-based" part of the model being low rank, not unlike our previous work \cite{scalable-physics-gp}. We assume $\sigma_n$ is known and estimate $V$ using the randomized sketching techniques where we are only required to evaluate $K_U$-vector products. To ensure a fair comparison we consider a sparse spiked model with $O(\log{n})$ spikes, which is equivalent to sketching $O(\log{n})$ low-rank components. In this case it requires $O(\log{n})$ $K_U$-vector products to estimate the low-rank components $V$ and the log-likelihood computation takes $O(n\log^2{n})$ time, which are both comparable to our HODLR model. We then compute the derivatives of the covariance matrix with respect to the parameters using the forward finite difference method since implementing the exact calculations would be a complex development effort. The score equations and the Fisher information matrix are thus computed using the numerical derivatives. Specifically we take the number of spikes to grow as $57\log_2(n)$, which was chosen so that at the lowest value of the data size we choose the computation would be exact. That is for the lowest $n$ considered  we have $57\log_2(n) \approx n$ (it is actually larger by 1 and we truncate it to $n$) so $K_U$ would be in that case the exact covariance matrix. For the HODLR method, we used $k=128$ for the off-diagonal rank in this experiment.

We subsample our datasets to generate observations of size $2^r$ for $r$ ranging from $9$ to $12$.  Recall that our dataset is simulated using the true parameter values $\theta_{\mathrm{true}}=(0.7,1,0.3,0.5)$. We explore the approximation accuracy of our approach both at the MLE point $(\hat{\rho},\hat{\sigma_\phi},\hat{\sigma_\chi},\hat{l})$ and at a potential starting point of the optimization, which was chosen to be $\theta_{\mathrm{init}}=(0.5,0.5,0.5,0.5)$. For both points, we use the standard relative precision to measure the approximation accuracy. Note that when the log-likelihood is maximized exactly, the score equations should be exactly 0. In this case the relative precision is undefined. We also consider two scaling-free measures at the estimated MLE point from \cite{geoga2020scalable},
\begin{align}
    \eta_g := ||S(\theta) - \tilde{S}(\theta)||_{\mathcal{I}(\theta)^{-1}},
    \label{eq:score-error}
\end{align}
for the score equations where $S(\theta),\tilde{S}(\theta)$ are the exact and approximated score equations respectively. We also measure 
\begin{align}
    \eta_{\mathcal{I}} := \mathrm{tr}\left( (\mathcal{I}(\theta) - \tilde{\mathcal{I}}(\theta)) (\mathcal{I}(\theta)^{-1} - \tilde{\mathcal{I}}(\theta)^{-1}) \right)^{1/2},
    \label{eq:fisher-error}
\end{align}
for the Fisher information matrix, which is a natural metric for positive definite matrices. We use $\epsilon_{\tilde{L}},\epsilon_{\tilde{S}},\epsilon_{\tilde{\mathcal{I}}}$ to denote the relative precision of log-likelihood, score equations and Fisher information matrix respectively. The averaged relative precision of the log-likelihood, score equations, and Fisher information matrix over all the five independently simulated datasets is summarized in part (a) of Tables \ref{tb:Relative precision init} and \ref{tb:Relative precision MLE} for our HODLR model and in part (b) of Tables \ref{tb:Relative precision init} and \ref{tb:Relative precision MLE} for the sparse spiked covariance model.

\begin{table}[htb!]
\centering
\begin{subtable}{1\textwidth}
\centering
    \begin{tabular}{|c|c|c|c|c|c|} 
        \hline
        & $n=2^9$ & $n=2^{10}$ & $n=2^{11}$ & $n=2^{12}$ & $n=2^{13}$\\
        \hline
        $\epsilon_{\tilde{L}}$ & $-6.72$ & $-6.13$ & $-5.82$ & $-3.76$ & $-3.79$\\
        \hline
        $\epsilon_{\tilde{S}}$ & $-5.71$ & $-4.73$ & $-4.71$ & $-2.07$ & $-2.60$\\
        \hline
        $\epsilon_{\mathcal{I}}$ & $-7.80$ & $-6.07$ & $-5.66$ & $-1.96$ & $-1.85$\\
        \hline
    \end{tabular}
    \caption{HODLR covariance estimation}
\end{subtable}\\
\begin{subtable}{1\textwidth}
\centering
    \begin{tabular}{|c|c|c|c|c|c|} 
        \hline
        & $n=2^9$ & $n=2^{10}$ & $n=2^{11}$ & $n=2^{12}$ & $n=2^{13}$\\
        \hline
        $\epsilon_{\tilde{L}}$ & $-15.58$ & $-2.15$ & $-1.07$ & $-1.01$ & $-0.71$\\
        \hline
        $\epsilon_{\tilde{S}}$ & $-7.42$ & $-1.22$ & $-0.36$ & $-0.24$ & $-0.09$\\
        \hline
        $\epsilon_{\mathcal{I}}$ & $-0.70$ & $-0.81$ & $-0.42$ & $-0.36$ & $-0.16$\\
        \hline
    \end{tabular}
    \caption{Sparse spiked covariance estimation}
\end{subtable}
\caption{Averaged relative precision (on $\mathrm{\log_{10}}$ scale) of the log-likelihood, score equations and observed Fisher information matrix. All the results are averaged for five datasets and are evaluated at the initial point of optimization $\theta_{\mathrm{init}}=(0.5,0.5,0.5,0.5)$.}
\label{tb:Relative precision init}
\end{table}

\begin{table}[htb!]
\centering
\begin{subtable}{1\textwidth}
\centering
    \begin{tabular}{|c|c|c|c|c|} 
        \hline
        & $n=2^9$ & $n=2^{10}$ & $n=2^{11}$ & $n=2^{12}$\\
        \hline
        $\epsilon_{\tilde{L}}$ & $-6.68$ & $-5.40$ & $-6.15$ & $-3.00$\\
        \hline
        $\epsilon_{\mathcal{I}}$ & $-6.47$ & $-5.16$ & $-5.90$ & $-1.83$\\
        \hline
        $\eta_g$ & $-8.29$ & $-5.15$ & $-5.57$ & $-0.58$\\
        \hline
        $\eta_{\mathcal{I}}$ & $-5.57$ & $-4.53$ & $-4.86$ & $-0.68$\\
        \hline
    \end{tabular}
    \caption{HODLR covariance estimation}
\end{subtable}\\
\begin{subtable}{1\textwidth}
\centering
    \begin{tabular}{|c|c|c|c|c|} 
        \hline
        & $n=2^9$ & $n=2^{10}$ & $n=2^{11}$ & $n=2^{12}$\\
        \hline
        $\epsilon_{\tilde{L}}$ & $-15.57$ & $-2.16$ & $-1.04$ & $-1.00$\\
        \hline
        $\epsilon_{\mathcal{I}}$ & $-0.77$ & $-0.59$ & $-0.41$ & $-0.33$\\
        \hline
        $\eta_g$ & $-11.02$ & $0.56$ & $1.99$ & $2.28$\\
        \hline
        $\eta_{\mathcal{I}}$ & $0.47$ & $0.22$ & $0.53$ & $0.64$\\
        \hline
    \end{tabular}
    \caption{Sparse spiked covariance estimation}
\end{subtable}
\caption{Averaged relative precision (on $\mathrm{\log_{10}}$ scale) of the log-likelihood, score equations and observed Fisher information matrix. All the results are averaged for five datasets and are evaluated at the MLE point. Here $\epsilon_{\tilde{S}}$ is removed since the exact score equations tend to zero at the MLE point.}
\label{tb:Relative precision MLE}
\end{table}

For the HODLR method, we see that the log-likelihood can be approximated very accurately, better than $0.1\%$ relative accuracy both at the starting point and the MLE point. The score equations have relative errors less than $1 \%$ at the starting point and significantly better for the smaller cases. The observed Fisher information has relative error less than $1.5\%$ (except on the scaling-free metrics on the largest case), and most times, significantly better than that,  for a quantity that most times goes to infinity for increased problems size. The accuracy is worse compared to the HODLR approach in \cite{geoga2020scalable}, though broadly comparable; we note, moreover, that that approach used explicit kernels. 

Notice that when $n=2^9$, $57\log_2(n)=513$ which we truncate to $512=2^9$. In that case the sparse spiked model is full-rank for the size of the problem. In that circumstance the log-likelihood can be estimated up to machine precision and the finite difference estimated score equations achieve square root of the machine precision. However the precision decays very quickly as the problem size grows, especially for the score equations and the Fisher information matrix. The error becomes even larger as the parameters getting closer to the exact MLE point estimates. Gradient-based optimization methods can fail with the unreliable gradient estimates. Other than $n=2^9$, the error of this approach compared to our HODLR method is worse by at least one order of magnitude for all metrics and on average much worse than that. While some of this may conceivably be due to the numerical approximation of the derivatives, that concern is inapplicable for the likelihood approximation where errors are worse by two orders of magnitude. We conclude that HODLR is significantly superior in accuracy to the sparse spiked model, and that its additional development complexity and effort is justified. 

\subsubsection{Parameter Estimation}

\begin{figure}[htb!]
\centering
 \includegraphics[width=1\textwidth]{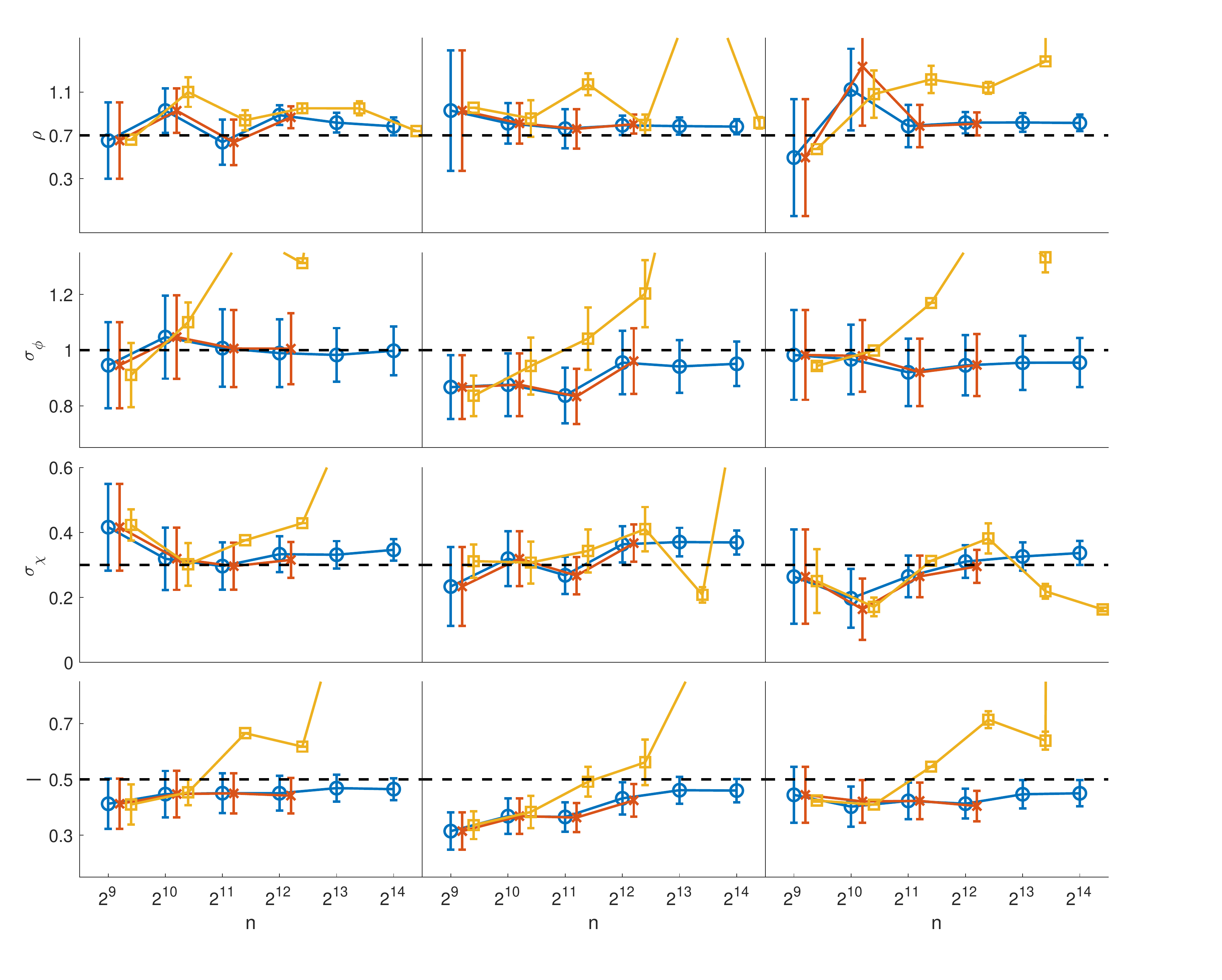}
\caption{Estimated MLEs and their $95\%$ confidence intervals using $n=2^r$ observations with $r$ from $8$ to $13$. Three columns in the figure represent results for three independently simulated datasets. The true parameter values $\theta_{\mathrm{true}}=(\rho,\sigma_\phi,\sigma_\chi,l)=(0.7,1,0.3,0.5)$ are added into each plot as black horizontal lines. We compare three models: (1) proposed HODLR model (blue curve with circles), (2) sparse spiked model (yellow curve with squares), (3) exact estimates (red line with x's) provided only for $r=8,\cdots,10$.}
\label{fig:parameter estimation}
\end{figure}

The previous subsections demonstrate both the scaling and the numerical accuracy of our proposed approximations. Both the two factors indicate their suitability for parameter estimations and uncertainty quantifications. Now we fit successively larger observations to obtain both the point estimates of the unknown parameters and their estimated confidence intervals using the observed Fisher information matrix. We take $n=2^r$ observations with $r$ from $9$ to $14$. In our experiments the HODLR off-diagonal rank is fixed at $k=128$. To solve the maximum likelihood estimation we use MATLAB's \texttt{fminunc} function with the default \texttt{quasi\_newton} algorithm to solve the unconstrainted optimization problem with respect to the parameters. We compare the exact model, the sparse spiked model and our proposed HODLR model. The initial guess is set to be $\theta_{\mathrm{init}}=(0.5,0.5,0.5,0.5)$. The stopping condition is chosen to be a relative tolerance of $10^{-6}$. For small sets of observations $r\leq 12$, we also provide parameter values estimated using the exact score equations. \figref{fig:parameter estimation} summarizes both the estimated parameters and their $95\%$ confidence intervals for three independently simulated datasets. 

As we can observe, the parameter estimates by our proposed approximations and the exact estimates are fairly close, both in terms of the parameter point estimates and the width of confidence intervals. Moreover, importantly, the true parameters are in the confidence intervals at our approach or very close to them except for large cases for the scale parameter, an issue also observed in \cite{geoga2020scalable}. We note that, for the Mat\'ern models, despite their ubiquity in spatial statistics, it is known that some parameters may not always be estimable \cite{zhang2004inconsistent,zhang2005towards,stein-theory-of-kriging}. Therefore, such an outcome, where the parameters are outside the confidence intervals, is possible. What needs to be observed, however, is that whenever the exact confidence intervals could be computed they were essentially the same as the HODLR approach, except that HODLR produced them much faster. In contrast, the sparse spiked model provides decent parameter point estimates and confidence interval estimations for problems of small scales ($n \leq 2^{10}$ for example). But the results quickly diverge as we move to large-scale dataset. The confidence interval estimations become even more problematic as we have seen in Table \ref{tb:Relative precision MLE}. In practice we can encounter indefinite Fisher information approximations, generating imaginary confidence interval width. In \figref{fig:parameter estimation} we only plot the real part of the confidence intervals. We conclude that the HODLR approach systematically produces confidence intervals comparable to the exact method, whereas the sparse spiked models has a variable and, sometimes, much worse accuracy than either.

We note that finding a good initial point for parameter optimizations can be difficult in practical applications. On the other hand, smaller off-diagonal ranks $k=36,72$ can also provide satisfying estimates with significantly faster runtime per iteration (since the complexity is at least quadratic with local rank $k$). A useful strategy is to try out different starting points with smaller off-diagonal rank $k$ and switch to larger rank when getting close to the MLE point. 

\subsection{Stationary Advection-diffusion-reaction Equation}

In this example, we consider a 2D advection-diffusion-reaction equation given by:
\begin{align}
    -\mathrm{div}(\kappa \nabla u) + \mathbf{v}\cdot \nabla u + cu &= \phi\quad\mathrm{in}\ \Omega=[-5,5]^2,
    \label{eq:ad1}\\
    \frac{\partial u}{\partial n} &= 0\quad \ \mathrm{on}\ \partial \Omega.
    \label{eq:ad2}
\end{align}
The coefficients $\kappa,\mathbf{v},c$ represent the diffusion, the advective velocity and the reaction constant, respectively. Here $\phi$ is the latent source term we are interested in and $u$ is the physical quantity that we can take measurements. Assume the source term $\phi$ is a Gaussian random field with Mat\'ern covariance function:
\begin{align}
    \phi\sim \mathcal{N}\left(20\cdot\mathrm{exp\left( -\frac{||\mathbf{x}||^2}{2\cdot 2^2} \right)} ,\sigma_{\phi}^2 M_{\nu,l}\right).
\end{align}
The mean function is known. Similarly by the SPDE representation, we can use \eqref{eq:latent covariance} to approximate the covariance. We use $L$ to denote the discretized elliptic operator induced by \eqref{eq:ad1} and \eqref{eq:ad2}. Then $Lu=\phi$ represents the elliptic PDE and $u$ has covariance model $K_u = \sigma_n I_n+ \Phi L^{-1}M_{\nu,l}L^{-T} \hat{\Phi}$, where $\sigma_n$ is again observation noise. We form the sparse matrix $L$ and use the backslash operator, effectively the standard LU factorization in Matlab.  The covariance matrix-vector product is done sequentially, and involves two PDE solves and one matrix-vector product with the Mat\'ern covariance SPDE approximation \eqref{eq:latent covariance}. 

For the numerical experiments below, the magnitude parameter $\sigma_{\phi}$ and length scale parameter $l$ are treated as the unknown parameters. We fix the smoothness parameter $\nu=1$, and choose the physical parameters $\kappa=0.001,c=0.5,\mathbf{v}=(x_1+5,x_2+5)$. To solve the PDE \eqref{eq:ad1} we discretize the domain $\Omega=[-5,5]^2$ using a finite element mesh. The approximation of the Mat\'ern covariance \eqref{eq:latent covariance} is also obtained via finite element computations. However it is performed on a slightly larger domain $[-7.5,7.5]^2$ to avoid the boundary effect.

The ground truth parameter values are taken to be $(\sigma_\phi,l)=(1,1)$. We use the true parameters to simulate five Mat\'ern random field using the \textit{R} package \textit{RandomFields} \cite{R-random-field}. Each dataset contains $2^{20}$ samples of $\phi$ on a regular 2D grid over $[-7.5,7.5]^2$. The observations of $u$ are then generated by solving the PDE \eqref{eq:ad1} and adding simulated measurement noise with $\sigma_n$ chosen so that the sample standard deviation is $20\%$ of the sample one. 

To approximate the covariance matrix to HODLR we use leaf level blocks with size between $256$ and $512$.  The maximum level of the HODLR approximation is thus $\left \lfloor{\log_2{(n/256)}}\right \rfloor $ where $n$ is both the size of $K_u$ and number of observations of $u$ (in other words, we choose $m=n$ here). For all off-diagonal blocks, a fixed rank $k=128$ is used. 

\subsubsection{Numerical Accuracy of the Log-likelihood, Score Equations and the Fisher Information}
\label{sec:accuracy-diff}

We demonstrate the numerical accuracy of the approximations for different sizes of observations. Assume the starting point of parameter estimation is $(\sigma_\phi,l)_{\text{init}}=(1.5,1.5)$.

Similarly we evaluate the accuracy using relative precision at the initial point and additionally measure \eqref{eq:score-error} and \eqref{eq:fisher-error} around the MLE point of the parameter estimations. All results in Table \ref{tb:Relative precision init diff} and Table \ref{tb:Relative precision MLE diff} are averaged from five independent datasets. We estimate the unknown parameters $(\sigma_\phi,l)$ by solving the approximated score equations \eqref{eq:score approximated}. \figref{fig:parameter estimation diffusion} illustrates both the estimated parameters and their $95\%$ confidence intervals for three independently simulated datasets. Similarly the optimization is conducted using the MATLAB's \texttt{fminunc} function with the default \texttt{quasi\_newton} algorithm with a relative tolerance of $10^{-6}$.  

\begin{table}[htb!]
\centering
\begin{tabular}{|c|c|c|c|c|c|} 
     \hline
     & $n=2^9$ & $n=2^{10}$ & $n=2^{11}$ & $n=2^{12}$ & $n=2^{13}$\\
     \hline
     $\epsilon_{\tilde{L}}$ & $-6.52$ & $-5.62$ & $-5.86$ & $-5.26$ & $-5.35$\\
     \hline
     $\epsilon_{\tilde{S}}$ & $-4.25$ & $-2.76$ & $-2.68$ & $-2.06$ & $-2.21$\\
     \hline
     $\epsilon_{\mathcal{I}}$ & $-6.10$ & $-5.44$ & $-4.96$ & $-3.88$ & $-2.47$\\
     \hline
\end{tabular}
\caption{Averaged relative precision (on $\mathrm{\log_{10}}$ scale) of the log-likelihood, score equations and observed Fisher information matrix. All the results are averaged for five datasets and are evaluated at the initial point of optimization $(\sigma_\phi,l)_{\text{init}}=(1.5,1.5)$.}
\label{tb:Relative precision init diff}
\end{table}

\begin{table}[htb!]
\centering
\begin{tabular}{|c|c|c|c|c|} 
 \hline
 & $n=2^{9}$ & $n=2^{10}$ & $n=2^{11}$ & $n=2^{12}$\\
 \hline
 $\epsilon_{\tilde{L}}$ & $-6.64$ & $-5.87$ & $-6.04$ & $-5.54$\\
 \hline
 $\epsilon_{\mathcal{I}}$ & $-6.41$ & $-5.60$ & $-5.38$ &  $-4.64$\\
 \hline
 $\eta_g$ & $-7.21$ & $-5.78$ & $-5.34$ & $-3.89$\\
 \hline
 $\eta_{\mathcal{I}}$ &  $-6.00$ & $-5.33$ & $-5.02$  & $-4.31$\\
 \hline
\end{tabular}
\caption{Averaged relative precision (on $\mathrm{\log_{10}}$ scale) of the log-likelihood, score equations and observed Fisher information matrix. All the results are averaged for five datasets and are evaluated at the MLE point. Here $\epsilon_{\tilde{S}}$ is removed since the exact score equations tend to zero at the MLE point.}
\label{tb:Relative precision MLE diff}
\end{table}



\subsubsection{Parameter Estimation}

\begin{figure}[htb!]
\centering
 \includegraphics[width=1\textwidth]{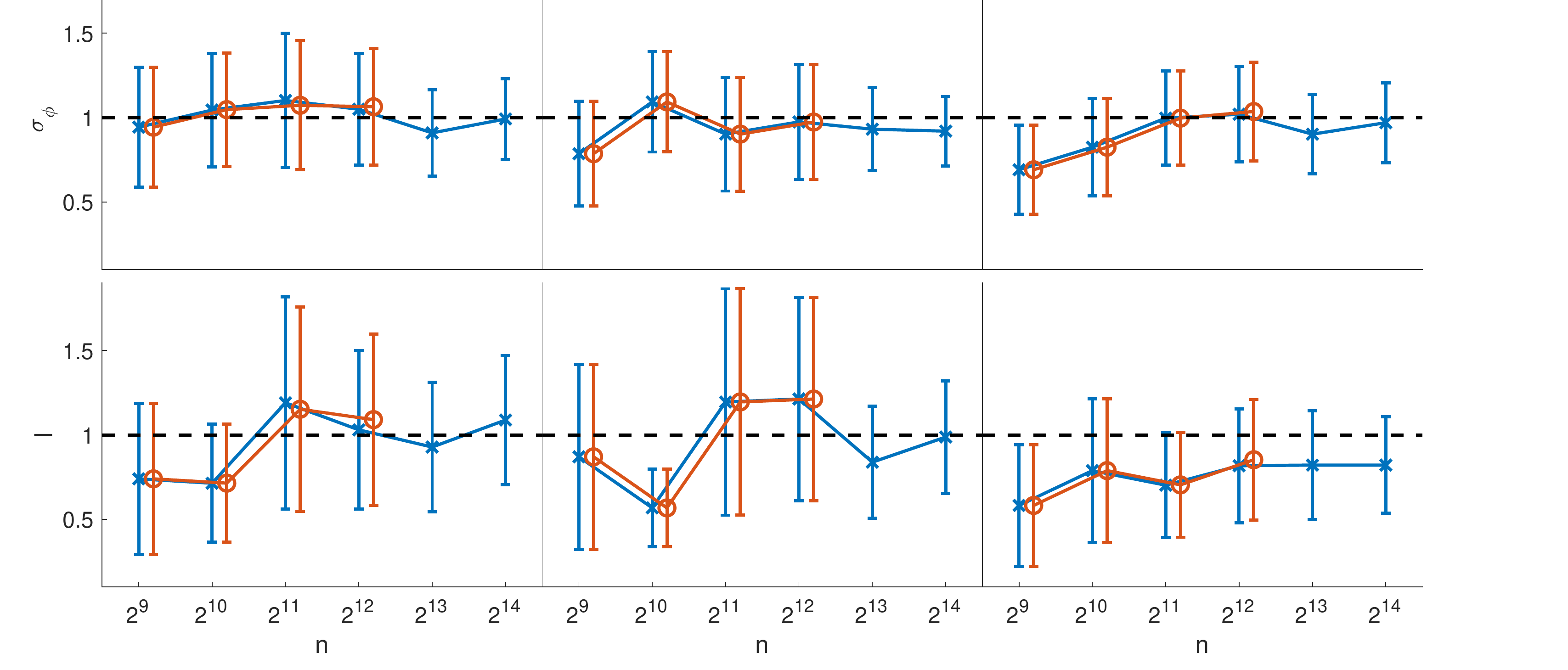}
 \caption{Estimated MLEs and their $95\%$ confidence intervals using $n=2^r$ observations with $r$ from $8$ to $13$. Three columns in the figure represent results for three independently simulated datasets. The true parameter values $\theta_{\mathrm{true}}=(\sigma_\phi,l)=(1,1)$ are added into each plot as black horizontal lines. Exact estimates (circle) are provided for $r=8,\cdots,10$.}
\label{fig:parameter estimation diffusion}
\end{figure}

The more relevant outcome, whether the true parameters are in the confidence intervals of the estimates obtained after the approximation, are however, better than those of \S \ref{sec:wind} particularly for large examples. The parameter point estimations and the confidence intervals are summarized in \figref{fig:parameter estimation diffusion}. The parameter $l$ is comparable in significance, though not in value, and, being smaller here it may be in itself the reason for the better fit. The approach certainly shows reasonable accuracy and consistency in our view, particularly given the fact that the covariance model is implicit. We point out that, in this section in particular, the model for $u$ requires solving a partial differential equation to get, even if one would do an explicit treatment of the Mat\'ern kernel itself (which would have been an option for the model in \S \ref{sec:wind}). Therefore in this case, for this model, our approach appears to be the only scalable alternative. 

\section{Discussion}
\label{sec:discussion}

In this paper, we propose a novel scheme for applying the well-known hierarchical matrices to Gaussian process maximum likelihood parameter estimation problem. For many spatial statistical problems, nearby observations have smoothly varying correlations with the rest of the system. By carefully reordering the observations and exploiting the structure, the off-diagonal blocks of the corresponding covariance matrix usually have fast decaying spectrum, which enables the application of hierarchical matrices to approximate the covariance matrix. This technique makes it possible to work with large datasets of measurements observed on unstructured grids. Estimations of the likelihood function, the score equations, and the expected Fisher information matrix all scale quasilinearly. Then the parameter estimations and uncertainty quantification can be obtained based on solving the score equation system and inverting the expected Fisher information matrix. Moreover, our approach uses ideas from \cite{fast-construct-of-hierarchical-matrices} to construct the covariance matrix using only $O(\log(n))$ matrix-vector products. For statistical models defined implicitly, by means of stochastic differential equations that allow fast solvers for their deterministic counterparts, this allows for a very efficient way of building the HODLR approximation while allowing for a very flexible way of statistical modeling, such as, for example when dealing with nonstationarity for which explicit models are difficult to generate. We also note that in the process we proposed a way to compute exactly (without re-compression) traces of products of HODLR matrices which are important for both efficient evaluation of the score equations and computing the features of the Fisher Information matrix. 

We note that the quality of the approximation depends crucially on the ordering of the observations and the local ranks of off-diagonal blocks. The proposed approach is a fixed rank strategy. An adaptive rank extension of the algorithm is certainly an important future direction to investigate. This can be done, potentially, by applying the adaptive randomized range sketching algorithm \cite{tropp-find-the-structure-with-randomness} with an online posterior estimate of the approximation error via samples. The choice of local ranks $k$ itself provides a balance between complexity and accuracy. Generally the local ranks grow moderately as the problem size grows. It is also meaningful to increase the rank when the estimations get close to the optimality. Moreover, the approximation decreases in quality with increasing level order, it would be interesting to explore way of slightly increase the access to the covariance matrix and stabilize it -- perhaps by combining it with stochastic estimators. 

\section*{Acknowledgment}
We thank the two anonymous referees for their careful reading of the manuscript and the suggestions that have improved its quality. 
This material is based upon work
supported by the U.S. Department of Energy, Office of Science,
Office of Advanced Scientific Computing Research (ASCR) under
Contract DE-AC02-06CH11357. 

\bibliographystyle{siamplain}
\bibliography{references}

\clearpage

\appendix

\section{Motivating Example for Full Dataset Fitting}\label{sec:motivating}

Before moving forward to sophisticated covariance and physical models, we demonstrate the benefit of performing MLE over the entire large datasets of measurements. Here we use a nonstationary Gaussian process on the 1$D$ interval $[0,10]$. The covariance model of the process is the square exponential covariance
\begin{align}
    K(x_i, x_j) = \sigma \exp{\left(-\frac{(x_i - x_j)^2}{2l_i l_j}\right)},\ \text{where }x_i,\ x_j\in [0,10]
\end{align}
where the length scale parameter $l_i$, $l_j$ are given by a linear function of $x$, i.e. $l_i=l(x_i)=a+bx_i$. We set the ground truth parameters $\sigma=1$, $a=0.1$, $b=0.6$. Therefore the true length scale parameter $l$ ranges from $0.1$ to $6.1$ over the entire domain. Here we treat the magnitude parameter $\sigma$ as known. By simulating synthetic observations from the given Gaussian process with an equally spaced grid points of mesh size $\Delta x=0.05$, we attempt to recover the unknown parameters $a$ and $b$ via MLE. 

We consider the following three settings:
\begin{itemize}
    \item Setting 1: full observations from $[0,10]$ with resolution $\Delta x=0.05$.
    \item Setting 2: subsampled observations from $[0,10]$ with a coarser resolution $\Delta x=0.2$.
    \item Setting 3: truncated observations from $[0, 2.5]$ with full resolution $\Delta x=0.05$.
\end{itemize}

The three settings corresponds to three scenarios when dealing with large datasets of raw observations. In setting 1, we perform MLE on the entire dataset and possibly apply approximation methods to confront the scaling issue. Alternatively we can subsample the dataset by taking a coarser subset or restricting the observations to a partial domain. These approaches will effectively reduce the problem size, but as a result, inducing difficulties and inaccuracy in parameter identifications, as what we will see next. 

\begin{figure}[htb!]
\centering
\includegraphics[width=0.4\textwidth]{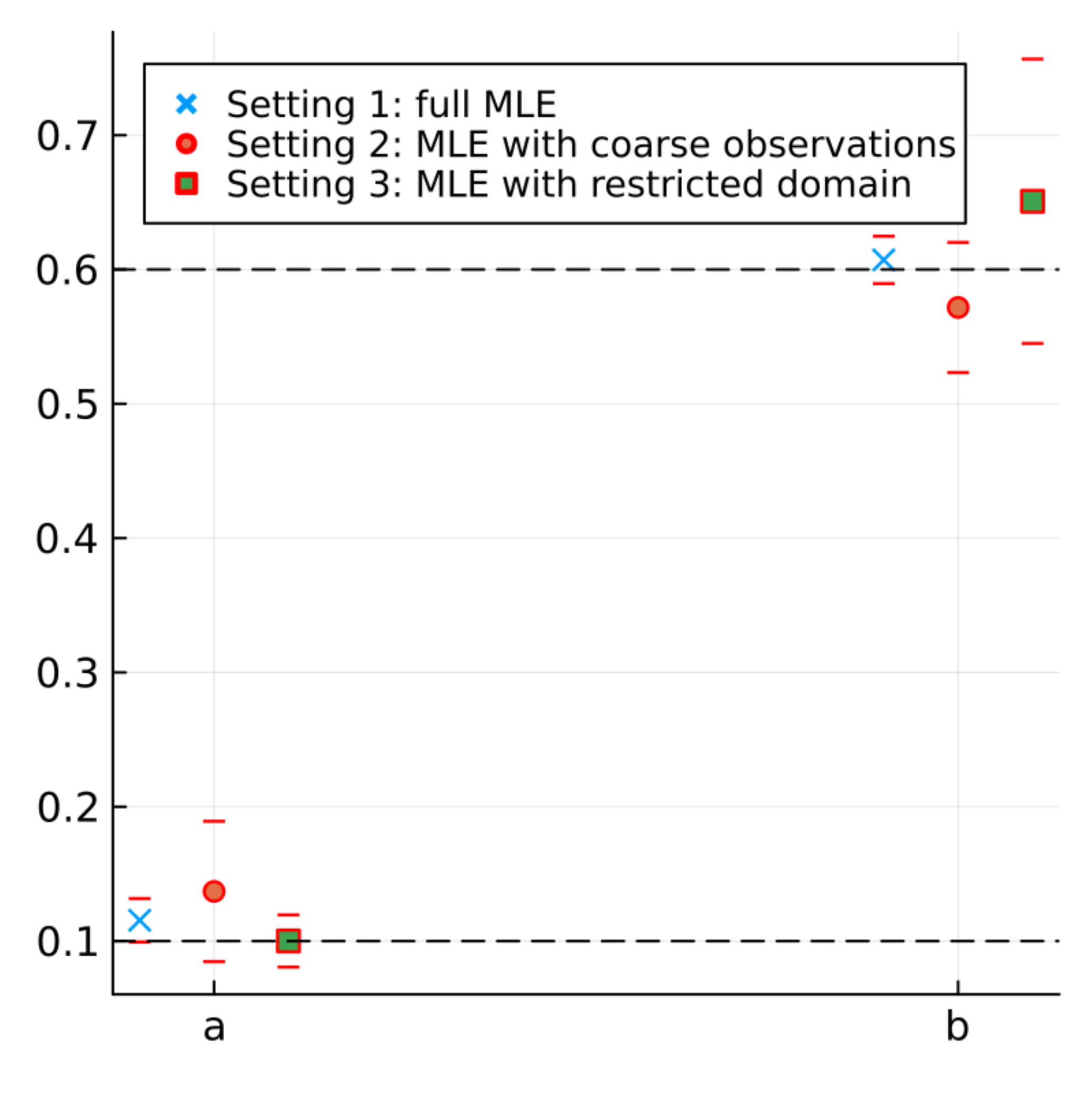}
\caption{MLE point estimation and confidence intervals for parameters $a$ and $b$. Blue crosses, red circles and green squares correspond to settings 1, 2, 3, respectively.}
\label{fig:toy_mle}
\end{figure}

We simulate $20$ independent realizations of the same process as observations and perform MLE for the three settings above. We show the parameter point estimations and the corresponding $95\%$ confidence intervals in \figref{fig:toy_mle}. We observe that with full observations we can identify both parameters accurately (with small confidence intervals). When we subsample the observations as in setting 2, the observation density ($\Delta x = 0.2$) is not enough to capture the finest correlation scale ($l_{\text{min}}=0.1$). We can still successfully identify the true parameter but the confidence intervals are much wider. In setting $3$, the truncated subdomain contains information about finer scale correlations, as the correlation length scale increases from $0$ to $10$. We see the intercept term $a$ can be estimated accurately but estimating $b$ poses much more uncertainties. In summary, it is indeed possible to subsample the observations to circumvent the scaling issue when dealing with large datasets. However this comes with a cost as partial observations may fail to provide enough information to identify the parameters. A better alternative is to use approximation method e.g. our approach in this paper, which introduces a clearer framework for complexity / accuracy tradeoff.

\section{Two Quasilinear Trace Operations for HODLR Matrices}
\label{app:two new operations}

 Assume $A$ and $B$ are both HODLR matrices with number of levels $\tau$ and fixed local rank $k$. Based on the basic factorization \eqref{eq:one way factor} we have
\begin{align}
    A = \bar{A}(I+U^{(\tau)}V^{(\tau)T})\cdots(I+U^{(1)}V^{(1)T}),
    \label{eq:factor 1}
\end{align}
where $\bar{A}$ is a block-diagonal matrix containing all the leaf level blocks of $A$. $(I+U^{(i)}V^{(i)T})$ is a block-diagonal matrix with $2^{i-1}$ blocks where each block is a rank-$2k$ update to identity. More details and the connection between off-diagonal blocks and low-rank update to identity can be found in Appendix \ref{app:hodlr-factorization}.

We assume the level grows as $\log{n}$, i.e. $\tau=O(\log{n})$. \eqref{eq:factor 1} is the original form of factorization in \cite{fast-direct-method-for-gaussian-process}. By factorizing the finer level diagonal factors to the other side, $A$ can also be factorized in the following ``transposed'' form in the same $O(n\log^2{n})$ complexity,
\begin{align}
    A = (I+V^{(1)}U^{(1)T})\cdots(I+V^{(\tau)}U^{(\tau)T})\bar{A}.
    \label{eq:factor 2}
\end{align}
Note that here we assume $A$ is symmetric. For asymmetric matrix the two factorizations \eqref{eq:factor 1}, \eqref{eq:factor 2} also exist, but the factors generally do not have explicit correspondence.

\subsection{Matrix-Matrix Product \texorpdfstring{$AB$}{AB} And \texorpdfstring{$A^{-1}B$}{A-1B}}
\label{app:AB,A^-1B}

Now we discuss in detail the algorithm for performing matrix-matrix product $AB$ and $A^{-1}B$ in HODLR format. Instead of forming the resulting product matrix explicitly, we write it as a sum of matrices. 
We start from HODLR matrices $A,B$ with the same number of levels $\tau$ and fixed local rank $k$. Additionally we assume $A,B$ have exactly the same size and hierarchical partitioning. Consider product $AB$. We use the factorization form \eqref{eq:factor 1}. 

\begin{itemize}
    \item[\textbf{Step 1}] The first step is to multiply the rightmost factor of $A$ with HODLR matrix $B$. We have
    \begin{align}
        (I+U^{(1)}V^{(1)T})B = B + U^{(1)}V^{(1)T}B = B + \bar{U}^{(1)}\bar{V}^{(1)T},
        \label{eq:tmp prod 2}
    \end{align}
    where $\bar{U}^{(1)}=U^{(1)}$ and $\bar{V}{(1)}=B^TV^{(1)}$. Note that the second term is of rank $k$. We can store the two low-rank factors $U_1^{(1)}$, $V_1^{(1)}$ to avoid computing it explicitly.
    
    The result now is written as the sum of an HODLR matrix and a low-rank component. The next step is to apply subsequent factors of $A$. This is done by two steps.
    
    \item[\textbf{Step 2}] Notice that the second level factor has the following block diagonal format  \cite{fast-direct-method-for-gaussian-process}, 
    \begin{align}
        (I+U^{(2)}V^{(2)T}) = 
        \begin{bmatrix}
            I+U_1^{(2)}V_1^{(2)T} & 0\\
            0 & I+U_2^{(2)}V_2^{(2)T}
        \end{bmatrix}.
        \label{eq:2nd level}
    \end{align}
   where each block has size one half of the entire matrix and the sized of the identity matrices is chose to match.   To apply the second level factor to the product, we first update the low-rank components. In this case we have
    \begin{align}
        \bar{U}^{(1)}\bar{V}^{(1)T}\xleftarrow{}(I+U^{(2)}V^{(2)T})\bar{U}^{(1)}\bar{V}^{(1)T},
    \end{align}
    which can be done by updating the existing low-rank factors $\bar{U}^{(1)}\xleftarrow{}(I+U^{(2)}V^{(2)T})\bar{U}^{(1)}$ and $\bar{V}^{(1)}\xleftarrow{}\bar{V}^{(1)}$ via applying the two diagonal blocks in \eqref{eq:2nd level} to the corresponding rows of $\bar{U}^{(1)}$.
    
    \item[\textbf{Step 3}] The next step is to apply the second level factor to the HODLR matrix. 
    Divide the HODLR matrix into its first level, $B$ can be written as
    \begin{align}
        B = 
        \begin{bmatrix}
            B_1^{(1)} & G_1^{(1)}H_1^{(1)T}\\
            H_2^{(1)}G_2^{(1)T} & B_2^{(1)}
        \end{bmatrix}.
    \end{align}
    Then we can compute
    \begin{align}
        (I+U^{(2)}V^{(2)T})B=
        &\begin{bmatrix}
            I+U_1^{(2)}V_1^{(2)T} & 0\\
            0 & I+U_2^{(2)}V_2^{(2)T}
        \end{bmatrix}
        \begin{bmatrix}
            B_1^{(1)} & G_1^{(1)}H_1^{(1)T}\\
            H_2^{(1)}G_2^{(1)T} & B_2^{(1)}
        \end{bmatrix},\nonumber\\
        =& 
        \begin{bmatrix}
            B_1^{(1)} & \bar{G}_1^{(1)}\bar{H}_1^{(1)T}\\
            \bar{H}_2^{(1)}\bar{G}_2^{(1)T} & B_2^{(1)}
        \end{bmatrix} + 
        \begin{bmatrix}
            U_1^{(2)}V_1^{(2)T}B_1^{(1)} & 0\\
            0 & U_2^{(2)}V_2^{(2)T}B_2^{(1)}
        \end{bmatrix}.
        \label{eq:tmp prod}
    \end{align}
    where $\bar{G}_1^{(1)}=(I+U_1^{(2)}V_1^{(2)T})G_1^{(1)}$, $\bar{H}_1^{(1)}=H_1^{(1)}$, $\bar{G}_2^{(1)}=G_2^{(1)}$, $\bar{H}_2^{(1)}=(I+U_2^{(2)}V_2^{(2)T})H_2^{(1)}$. Notice the first term is still an HODLR matrix and the second term is block-diagonally low-rank. Further denote
    \begin{align}
        \begin{bmatrix}
            U_1^{(2)}V_1^{(2)T}B_1^{(1)} & 0\\
            0 & U_2^{(2)}V_2^{(2)T}B_2^{(1)}
        \end{bmatrix} = 
        \begin{bmatrix}
            \bar{U}_1^{(2)}\bar{V}_1^{(2)T} & 0\\
            0 & \bar{U}_2^{(2)}\bar{V}_2^{(2)T}
        \end{bmatrix} = 
        \bar{U}^{(2)}\bar{V}^{(2)T},
        \label{eq:barV}
    \end{align}
    and define the HODLR matrix $\tilde{B}$ in the multiplier by
    \begin{align}
        \tilde{B} \xleftarrow[]{}
        \begin{bmatrix}
            B_1^{(1)} & \bar{G}_1^{(1)}\bar{H}_1^{(1)T}\\
            \bar{H}_2^{(1)}\bar{G}_2^{(1)T} & B_2^{(1)}
        \end{bmatrix}.
    \end{align}
Note that $B$ and $\tilde{B}$ are HODLR matrices with the same size and hierarchical structure. However the off-diagonal blocks have been updated. Now the product can be written as
    \begin{align}
        (I+U^{(2)}V^{(2)T})(I+U^{(1)}V^{(1)T})B=\tilde{B} + \bar{U}^{(1)}\bar{V}^{(1)T} + \bar{U}^{(2)}\bar{V}^{(2)T}.
        \label{eq:tmp prod 3}
    \end{align}
    We notice not only the HODLR matrix has been updated, an extra second level diagonal low-rank component ($\bar{U}^{(2)}\bar{V}^{(2)T}$ in our case) has been generated as well.
    
    \item[\textbf{Step 4}] Steps 2 through 4 are repeated until we reach level $\tau$. For each level, both the HODLR matrix $\bar{B}$ and all the low-rank components from the previous levels need to be updated. Finally we have
    \begin{align}
        (I+U^{(\tau)}V^{(\tau)T})\cdots(I+U^{(1)}V^{(1)T})B=\tilde{B} + \sum_{i=1}^{\tau}\bar{U}^{(i)}\bar{V}^{(i)T}.
        \label{eq:AB result}
    \end{align}
  Here $\tilde{B}$ is a HODLR matrix with the same structure as $B$. Note that each $\bar{U}^{(i)}$ and $\bar{V}^{(i)}$ have the same dimension and structure as $U^{(i)}$ and $V^{(i)}$  : they are block diagonal with $n/2^{i-1} \times 2k$  blocks.
    \item[\textbf{Step 5}] The final step is to apply the leaf level block diagonal matrix $\bar{A}$ to the product we have gotten. Since we assume $A,B$ have the same hierarchical partitioning, it can be done by applying each diagonal block of $\bar{A}$ blockwisely to all low-rank components and the low-rank factors of $\bar{B}$ at all levels. The operation has been extensively used when factorizing the HODLR matrix, see \cite{sym-factorization-of-hodlr} for an example.
\end{itemize}

Now we analyze the computational complexity of the given workflow. Using \eqref{eq:factor 1}, we apply the level $i$ factor of $A$. This requires multiplying the block-diagonal matrix $(I+U^{(i)}V^{(i)T})$ with all $(i-1)$ existing left block-diagonally low-rank components $\bar{U}^{(1)},\cdots,\bar{U}^{(i-1)}$.  \textit{The essential observation that makes this efficient is that block diagonal structures at level $i$ can be mapped into block diagonal structures at all other levels $k$, $k \leq i$ since they have the finest structures among the latter.} Take $\bar{U}^{(i-1)}$ as an example, exploring the preceding observation, we can express level $i$ in block diagonal form for the $i-1$ partition and compute the matrix multiplication as follows,

\begin{align}
    \label{eq:(I+low-rank)*low-rank}
    & (I+U^{(i)}V^{(i)T}) \bar{U}^{(i-1)}, \\
    =\ &
    \bdm{I+U_1^{(i)}V_1^{(i)T}}{\cdots}{I+U_{2^{i-1}}^{(i)}V_{2^{i-1}}^{(i)T}}
    \cdot
    \bdm{\bar{U}_1^{(i-1)}}{\cdots}{\bar{U}_{2^{i-2}}^{(i-1)}},\nonumber\\
    =\ & \bdm{\left(I+U_1^{(i)}V_1^{(i)T}\right) \bar{U}_{1,1}^{(i-1)}}{\left(I+U_2^{(i)}V_2^{(i)T}\right) \bar{U}_{1,2}^{(i-1)}}{\cdots}{\left(I+U_{2^{i-1}}^{(i)}V_{2^{i-1}}^{(i)T}\right) \bar{U}_{2^{i-2},2}^{(i-1)}}.\nonumber
\end{align}
where $\mathrm{diag}(v)$ denotes the block-diagonal matrix whose diagonal blocks are given by the component block matrices $v$. $\bar{U}_1^{(i-1)}=\begin{bmatrix}\bar{U}_{1,1}^{(i-1)}\\\bar{U}_{1,2}^{(i-1)}\end{bmatrix}$ is a partition of matrix $\bar{U}_1^{(i-1)}$ compatible with the finer partition $i$. 
To evaluate the complexity consequences of our approach, we need to investigate the size of the block matrices in detail. For simplicity we assume bipartition for all levels in the hierarchical partitioning structure and therefore that $n$ is divisible by $2^\tau$. In this case, $(I+U^{(i)}V^{(i)T})$ has diagonal blocks of size $n/2^{i-1}\times n/2^{i-1}$. In contrast, $\bar{U}^{(i-1)}$ has diagonal blocks of size $n/2^{i-2}\times 2k$. 
Therefore $\bar{U}_{1,1}^{(i-1)},\bar{U}_{1,2}^{(i-1)}$ are both of size $n/2^{i-1}\times 2k$. To solve \eqref{eq:(I+low-rank)*low-rank}, we need to compute matrix multiplications of type $\left(I+U_1^{(i)}V_1^{(i)T}\right) \bar{U}_{1,1}^{(i-1)}$ for all $2^{i-1}$ blocks. By multiplying the low-rank factors from right to left, each term takes $8nk^2/2^{i-1}$, yielding a total cost of $8nk^2$. At the following level, $\bar{U}^{(i-2)}$ has diagonal blocks of size $n/2^{i-3}\times 2k$. Therefore each diagonal block of of  should be quartered to match the multiplier $(I+U^{(i)}V^{(i)T})$. Ultimately there are still $2^{i-1}$ block multiplications of the same size as $\left(I+U_1^{(i)}V_1^{(i)T}\right) \bar{U}_{1,1}^{(i-1)}$. The computational cost is $8nk^2$ as well. Repeating the updating procedure for all $(i-1)$ existing low-rank components, the total cost is $8(i-1)nk^2$.

Next, updating $\tilde{B}$ requires applying $(I+U^{(i)}V^{(i)T})$, $i \geq 2$, which has $2^{i-1}$ blocks \eqref{eq:tr(A^-1B) computation},   to each of the left low-rank off-diagonal components of $\tilde{B}$ at levels $1,\cdots,i-1$, \eqref{eq:tmp prod} .  The same trick detailed in \eqref{eq:(I+low-rank)*low-rank} can be applied as well. For example in level $1$, two low-rank factors of size $n/2\times k$ need to be updated ($\bar{G}_1^{(1)}$ and $\bar{H}_2^{(1)}$ in \eqref{eq:tmp prod}). By partitioning each low rank component of $\tilde{B}$ into $2^{i-2}$ blocks (since $i=2$ no splitting was required in \eqref{eq:tmp prod}), we can match them with the diagonal blocks in $(I+U^{(i)}V^{(i)T})$. The resulting $2^i$ block multiplications take $4nk^2$ time. For all $(i-1)$ levels, the cost is $4(i-1)nk^2$.

Additionally to generate the new low-rank terms, $\bar{U}^{(i)}$, $\bar{V}^{(i)}$ in \eqref{eq:tmp prod 3} and \eqref{eq:AB result}, we need to apply the diagonal low-rank components of $(I+U^{(i)}V^{(i)T})$, i.e. $U_j^{(i)}V_j^{(i)T}$ in \eqref{eq:tr(A^-1B) computation} to the diagonal components of $\tilde{B}$ on level $i$ (i.e  the computation of \eqref{eq:barV}). This step can be done by regular HODLR-vector products (after transposing the computation). The complexity of HODLR-vector products has been extensively studied, see \cite{hierarchical-matrix-analysis}. Note that the diagonal component matrices of $\tilde{B}$ at level $i$ have size $n/2^{i}\times n/2^{i}$ and $(\tau-i)$ levels of their own. The total computational cost is thus upper bounded $O(nk^2\tau)$.

Adding all these operations, we obtain a total computational cost,

\begin{align}
    \sum_{i=1}^{\tau} 8(i-1)nk^2+4(i-1)nk^2+ O(nk^2\tau)=O(nk^2\tau^2)=O(n\log^2{n}).
\end{align}
Here we used the assumption $\tau=O(\log{n})$ and that $k$ is a fixed constant. Finally we need to apply all $2^\tau$ leaf blocks of $\bar{A}$ to $\tilde{B} + \sum_{i=1}^{\tau}\bar{U}^{(i)}\bar{V}^{(i)T}$. That is equivalent to applying all leaf blocks of $\bar{A}$ to all the left low-rank components and leaf blocks of $\tilde{B}$, and all $\bar{U}^{(i)}$. By blockwise application, each diagonal leaf block of $\bar{A}$ will be applied to $2k$ vectors (left low-rank components of off-diagonal blocks) in each level and an extra $O(k)$ vectors for the leaf blocks of $\tilde{B}$, yielding $O(k^3\tau)$ complexity. Similarly, to multiply with $\sum_{i=1}^{\tau}\bar{U}^{(i)}\bar{V}^{(i)T}$, each $\bar{A}$ leaf block needs to be applied to $2k$ vectors (diagonal blocks of $\bar{U}^{(i)}$), yielding $O(k^3\tau)$ complexity. For all blocks, the total complexity is $O(k^3\tau\times 2^\tau)=O(nk^2\tau)=O(n\log{n})$. Summarizing everything up, the total cost of computing $AB$ for two HODLR matrices in the format of the right hand side of \eqref{eq:AB result} is $O(n\log^2{n})$.

Next we discuss the algorithm for computing $A^{-1}B$ in a format of the right hand side of  \eqref{eq:AB result}. Since the inverse will reverse the order of the factors, we factorize $A$ in form \eqref{eq:factor 2}. Using the Woodbury identity, we have
\begin{align}
    A^{-1}B&=\bar{A}^{-1}(I+U^{(\tau)}V^{(\tau)T})^{-1}\cdots(I+U^{(1)}V^{(1)T})^{-1}B\nonumber\\
    &=\bar{A}^{-1}(I-U^{(\tau)}(I+V^{(\tau)T}U^{(\tau)})^{-1}V^{(\tau)T})\cdots (I-U^{(1)}(I+V^{(1)T}U^{(1)})^{-1}V^{(1)T})B.
    \label{eq:A^-1B goal}
\end{align}

Comparing \eqref{eq:AB result} and \eqref{eq:A^-1B goal}, each factor is still a low-rank update to identity. The only differences are now we need to apply the inverse of a matrix $(I+V^{(i)T}U^{(i)})$ and the inverse of leaf blocks of $\bar{A}$. Luckily, $V^{(i)T}U^{(i)}$ consists of only diagonal blocks of size $O(k)$ which makes the linear system efficiently computable. If we denote $\tilde{V}^{(i)T}=(I+V^{(i)T}U^{(i)})^{-1}V^{(i)T}$ we can apply the same algorithm we did to compute $AB$ to compute $A^{-1}B$ in the format of the right hand side of \eqref{eq:AB result}.

Now we analyze the additional computational complexity compared to the original $AB$ product algorithm. One of the extra operations is to obtain $\tilde{V}^{(i)T}=(I+V^{(i)T}U^{(i)})^{-1}V^{(i)T}$. Recall that both $U^{(i)}$ and $V^{(i)}$ are block-diagonal with a total number of $2^{i-1}$ diagonal blocks of rank $2k$. For each level $i$, the diagonal blocks of $V^{(i)}$ are of size $n/2^{i-1}\times 2k$. Therefore the complexity of computing each block of $\tilde{V}^{(i)}$ is $8nk^3/2^{i-1}$. For all blocks the complexity becomes $8nk^3$. Repeating the same operation to obtain all $\tilde{V}^{(i)}$, $i=1,\cdots,\tau$ requires $O(nk^3\tau)=O(n\log{n})$ complexity. Another difference is that now we need to apply the inverse of leaf blocks in $\bar{A}$. Based on similar considerations, each leaf block needs to be applied to $O(k\tau)$ vectors, yielding $O(k^4\tau)$ complexity. Taking all blocks into consideration, the total extra complexity is $O(k^4\tau\times 2^\tau)=O(nk^3\tau)=O(n\log{n})$.



In summary, the total cost of computing either $AB$ or $A^{-1}B$ for two HODLR matrices scales as $O(n\log^2{n})$ and the result can be expressed as the sum of an HODLR matrix and $\tau$ terms with low rank blocks, as in the right hand side of \eqref{eq:AB result}. The extra memory required to store the low-rank terms is $O(nk\tau)=O(n\log{n})$.

\subsection{Product of Form \texorpdfstring{$A^{-1}BC^{-1}D$}{A-1BC-1D}}
\label{app:A^-1BC^-1D}

In the same vein, we can compute the product of form $A^{-1}BC^{-1}D$ given four HODLR matrices $A,B,C,D$. Assume all four HODLR matrices have exactly the same size of hierarchical partitioning. The local rank of all off-diagonal blocks is fixed at $k$. Using \ref{app:AB,A^-1B} we are able to compute $A^{-1}B$ and $C^{-1}D$ separately. Assume 
\begin{align}
    A^{-1}B &= \bar{B} + \sum_{i=1}^{\tau}\bar{U}^{(i)}\bar{V}^{(i)T},
    \label{eq:A^-1B}\\
    C^{-1}D &= \bar{D} + \sum_{i=1}^{\tau}\bar{\mathbf{U}}^{(i)}\bar{\mathbf{V}}^{(i)T}.
\end{align}

By multiplying each term separately, we have
\begin{align}
    A^{-1}BC^{-1}D = \bar{B}\bar{D} + \sum_{i=1}^{\tau}\bar{U}^{(i)}\bar{V}^{(i)T}\bar{D} + \sum_{i=1}^{\tau}\bar{B}\bar{\mathbf{U}}^{(i)}\bar{\mathbf{V}}^{(i)T} + \sum_{i=1}^{\tau}\sum_{j=1}^{\tau}\bar{U}^{(i)}\bar{V}^{(i)T}\bar{\mathbf{U}}^{(j)}\bar{\mathbf{V}}^{(j)T}.
\end{align}

The first term is the product of two HODLR matrices. Applying the algorithms in \ref{app:AB,A^-1B} again, we can write
\begin{align}
    \bar{B}\bar{D} = \tilde{D} + \sum_{i=1}^{\tau}\bar{\mathcal{U}}^{(i)}\bar{\mathcal{V}}^{(i)T}.
\end{align}

In summary, the product can be written as
\begin{align}
    A^{-1}BC^{-1}D = \tilde{D} + \sum_{i=1}^{\tau}\bar{\mathcal{U}}^{(i)}\bar{\mathcal{V}}^{(i)T} + \sum_{i=1}^{\tau}\bar{U}^{(i)}\bar{V}^{(i)T}\bar{D} + \sum_{i=1}^{\tau}\bar{B}\bar{\mathbf{U}}^{(i)}\bar{\mathbf{V}}^{(i)T} + \sum_{i=1}^{\tau}\sum_{j=1}^{\tau}\bar{U}^{(i)}\bar{V}^{(i)T}\bar{\mathbf{U}}^{(j)}\bar{\mathbf{V}}^{(j)T}.
    \label{eq:A^-1BC^-1D}
\end{align}

From \ref{app:AB,A^-1B}, computing $A^{-1}B$, $C^{-1}D$, $\bar{B}\bar{D}$ all take $O(n\log^2{n})$ time. In total, given HODLR matrices $A,B,C,D$ we can write the product of $A^{-1}BC^{-1}D$ in form \eqref{eq:A^-1BC^-1D} with a cost of $O(n\log^2{n})$ complexity. Though still complicated, we will show how to combine \eqref{eq:A^-1BC^-1D} with the trace operation to speed up the computation.

\subsection{Computation of \texorpdfstring{$\mathrm{tr(A^{-1}B)}$}{tr(A-1B)} And \texorpdfstring{$\mathrm{tr(A^{-1}BC^{-1}D)}$}{tr(A-1BC-1D)}}
\label{app:trace}

Taking the trace of form \eqref{eq:A^-1B}, we have
\begin{align}
    \mathrm{tr}(A^{-1}B) &= \mathrm{tr(\bar{B})} + \sum_{i=1}^{\tau}\mathrm{tr}(\bar{U}^{(i)}\bar{V}^{(i)T}).
\end{align}

Further recall that $\bar{U}^{(i)}\bar{V}^{(i)T}$ is a block diagonal low-rank matrix, which can be written as
\begin{align}
    \bar{U}^{(i)}\bar{V}^{(i)T} = 
    \begin{bmatrix}
        \bar{U}_1^{(i)}\bar{V}_1^{(i)T} & 0 & \cdots & 0\\
        0 & \bar{U}_2^{(i)}\bar{V}_2^{(i)T} & \cdots & 0\\
        \vdots & \vdots & \ddots & 0\\
        0 & 0 & 0 & \bar{U}_{2^{i-1}}^{(i)}\bar{V}_{2^{i-1}}^{(i)T}
    \end{bmatrix}.
    \label{eq:tr(A^-1B) computation}
\end{align}

Now using the properties of the trace operator, we have
\begin{align}
    \mathrm{tr}(A^{-1}B) &= \mathrm{tr(\bar{B})} + \sum_{i=1}^{\tau}\mathrm{tr}(\bar{V}^{(i)T}\bar{U}^{(i)}),\\
    &=\mathrm{tr(\bar{B})} + \sum_{i=1}^{\tau}\sum_{j=1}^{2^i}\mathrm{tr}(\bar{V}_j^{(i)T}\bar{U}_j^{(i)}).
\end{align}

Notice that both $\bar{U}_j^{(i)},\bar{V}_j^{(i)}\in\mathbb{R}^{n/2^{i-1}\times 2k}$. Computing the product $\bar{V}_j^{(i)T}\bar{U}_j^{(i)}$ can be done in $(8nk^2/2^{i-1})$ time, extracting its trace requires $k-1$ additions, and it is asymptotically negligible. Repeating the calculation for all $2^{i-1}$ terms at all levels $i=1,\dots,\tau$, the total computational cost is $O(nk^2\tau)=O(n\log{n})$. Taking into account the cost of producing \eqref{eq:A^-1B}, the total computational complexity of computing $\mathrm{tr}(A^{-1}B)$ for two HODLR matrices is $O(n\log^2{n})$.

Next we consider taking the trace of \eqref{eq:A^-1BC^-1D}. Similarly utilizing the basic trace properties, we can write
\begin{align}
    \mathrm{tr}(A^{-1}BC^{-1}D) = &\mathrm{tr}(\tilde{D}) + \sum_{i=1}^{\tau}\sum_{j=1}^{2^i}\mathrm{tr}(\bar{\mathcal{V}}_j^{(i)T}\bar{\mathcal{U}}_j^{(i)}) + \sum_{i=1}^{\tau}\mathrm{tr}(\bar{V}^{(i)T}\bar{D}\bar{U}^{(i)})\nonumber\\
    &+ \sum_{i=1}^{\tau}\mathrm{tr}(\bar{\mathbf{V}}^{(i)T}\bar{B}\bar{\mathbf{U}}^{(i)}) + \sum_{i=1}^{\tau}\sum_{j=1}^{\tau}\mathrm{tr}(\bar{\mathbf{V}}^{(j)T}\bar{U}^{(i)}\bar{V}^{(i)T}\bar{\mathbf{U}}^{(j)}).
    \label{eq:tr(A^-1BC^-1D)}
\end{align}

Since $\tilde{D}$ is of HODLR form, taking its trace is very efficient and can be negligible. Next we consider the second term. Following the same procedure in \eqref{eq:tr(A^-1B) computation}, we can compute the second term with cost $O(n\log{n})$.

Moving to the third and fourth terms, they share the same structure. Take the third term $\sum_{i=1}^{\tau}\mathrm{tr}(\bar{V}^{(i)T}\bar{D}\bar{U}^{(i)})$ as an example. Note $\bar{D}$ is an HODLR matrix and $\bar{U}^{(i)}$, $\bar{V}^{(i)}$ are block-diagonal matrices with $2^{i-1}$ diagonal blocks of size $(n/2^{i-1})\times 2k$. Therefore the trace operations depends only on the diagonal blocks of $\bar{D}$ at level $i-1$ of size $(n/2^{i-1})\times (n/2^{i-1})$. There are $2^{i-1}$ of them in total, each having HODLR structure with $\tau-i+1$ levels:
\begin{align}
    \bar{D}_1^{(i-1)},\bar{D}_2^{(i-1)},\cdots,\bar{D}_{2^{i-1}}^{(i-1)}.
\end{align}
We then obtain
\begin{align}
    \mathrm{tr}(\bar{V}^{(i)T}\bar{D}\bar{U}^{(i)}) = \sum_{j=1}^{2^{i-1}} \mathrm{tr}(\bar{V}_j^{(i)T} \bar{D}_{j}^{(i-1)} \bar{U}_j^{(i)}).
\end{align}

The computations can be conducted efficiently by computing $\bar{D}_{j}^{(i-1)} \bar{U}_j^{(i)}$ first which can be done by HODLR-vector product. Similar operations have been analyzed in detail in \ref{app:AB,A^-1B}. The sum-total complexity for all $2^{i-1}$ terms is upper bounded by $O(nk^2\tau)$ and each product is of size $(n/2^{i-1})\times 2k$. Then we can left-multiply $\bar{V}_j^{(i)T}$ with the result via regular matrix-matrix multiplication. The sum-total complexity is $O(nk^2)$ for all $2^{i-1}$ terms. Each product is now of size $2k\times 2k$ and the complexity of the trace operation is $O(k 2^{i-1})$, which thus can be ignored. Repeating the same operations for all $\tau$ terms in $\sum_{i=1}^{\tau}\mathrm{tr}(\bar{V}^{(i)T}\bar{D}\bar{U}^{(i)})$, the total complexity is given by $O(nk^2\tau^2)=O(n\log^2{n})$. The fourth term $\sum_{i=1}^{\tau}\mathrm{tr}(\bar{\mathbf{V}}^{(i)T}\bar{B}\bar{\mathbf{U}}^{(i)})$ can be computed similarly with the same scaling.

For the last term, we compute $\bar{\mathbf{V}}^{(j)T}\bar{U}^{(i)}$ and $\bar{V}^{(i)T}\bar{\mathbf{U}}^{(j)}$ for each pair of $(i,j)$ separately. By applying blockwise matrix multiplications, both can be computed in $O(nk^2)$ time. Now we analyze their output matrices to conduct the following operations. 

If $i\geq j$, the product $\bar{\mathbf{V}}^{(j)T}\bar{U}^{(i)}$ is a block-diagonal matrix with $2^{j-1}$ diagonal blocks. Each diagonal block is of size $k\times 2^{(i-j)}k$. $\bar{V}^{(i)T}\bar{\mathbf{U}}^{(j)}$ is also a block-diagonal matrix with $2^{j-1}$ diagonal blocks. Its block has size $2^{(i-j)}k\times k$. In this case we compute their product directly by multiplying the corresponding diagonal blocks. Each pair of diagonal blocks takes $O(2^{(i-j)}k^3)$ time. In total, all pairs take $O(2^{i}k^3)$ time.

If $j\geq i$, we swap the order of the two matrices in the trace, i.e. $\mathrm{tr}(\bar{\mathbf{V}}^{(j)T}\bar{U}^{(i)}\bar{V}^{(i)T}\bar{\mathbf{U}}^{(j)}) = \mathrm{tr}(\bar{V}^{(i)T}\bar{\mathbf{U}}^{(j)}\bar{\mathbf{V}}^{(j)T}\bar{U}^{(i)})$. Now similarly, $\bar{V}^{(i)T}\bar{\mathbf{U}}^{(j)}$ is block-diagonal with $2^{i-1}$ blocks. Each block is of size $k\times 2^{(j-i)}k$. $\bar{\mathbf{V}}^{(j)T}\bar{U}^{(i)}$ is block-diagonal with $2^{i-1}$ blocks. Each block is of size $2^{(j-i)}k \times k$. Now we multiply all the pairs of diagonal blocks. The total cost is $O(2^{j}k^3)$.

Combining two cases together, the cost of computing the block-diagonal matrix inside the trace operator takes $O(2^{\mathrm{max}(i,j)}k^3)$ time. The complexity for the following trace operations can be ignored. In total, evaluating the last term takes
\begin{align}
    \sum_{i=1}^{\tau}\sum_{j=1}^{\tau} O(nk^2+k^3 2^{\mathrm{max}(i,j)}) = O(nk^2\tau^2+2^\tau k^3\tau) = O(n\log^2{n}).
\end{align}

In summary, the total cost of evaluating \eqref{eq:tr(A^-1BC^-1D)} is $O(n\log^2{n})$. \ref{app:AB,A^-1B}, \ref{app:A^-1BC^-1D}, \ref{app:trace} streamlined an exact approach of evaluating operations of form $\mathrm{tr(A^{-1}B)}$ and $\mathrm{tr(A^{-1}BC^{-1}D)}$ for HODLR matrices. Given the HODLR form, both operations take $O(n\log^2{n})$ time and an extra memory of $O(n\log{n})$.

\section{HODLR factorization}
\label{app:hodlr-factorization}

HODLR matrices admit several fast factorization algorithms. Particularly, factorization of form \eqref{eq:one way factor} is extensively used in this work. Here we use an example to explain the algorithm in detail and more importantly, establish a relationship between HODLR form \eqref{eq:hodlr format} and its block-diagonal factors in \eqref{eq:one way factor}. We use the same example of a $2$-level HODLR matrix as in \eqref{eq:hodlr format},

\begin{align}
    A = \begin{bmatrix}
    \begin{bmatrix}
    A_1^{(2)} & W_1^{(2)}X_1^{(2)T}\\
    X_1^{(2)}W_1^{(2)T} & A_2^{(2)}
    \end{bmatrix} & W_1^{(1)}X_1^{(1)T}\\
    X_1^{(1)}W_1^{(1)T} & 
    \begin{bmatrix}
    A_3^{(2)} & W_2^{(2)}X_2^{(2)T}\\
    X_2^{(2)}W_2^{(2)T} & A_4^{(2)}
    \end{bmatrix}
    \end{bmatrix},
\end{align}
where diagonal blocks $A_i^{(2)}$ in the leaf level (level 2) are assumed to be dense. The first step is to factor out the dense leaf blocks from $A$. For example we can factor out the leaf blocks from the left. Let us denote
\begin{align}
    \bar{A} = \begin{bmatrix}
    A_1^{(2)} & 0 & 0 & 0\\
    0 & A_2^{(2)} & 0 & 0\\
    0 & 0 & A_3^{(2)} & 0\\
    0 & 0 & 0 & A_4^{(2)}
    \end{bmatrix},
\end{align}
then we can compute blockwisely by
\begin{align}
    A &= 
    \bar{A}
    \begin{bmatrix}
    \begin{bmatrix}
    I_{n/4} & \left(A_1^{(2)}\right)^{-1} W_1^{(2)}X_1^{(2)T}\\
    \left(A_2^{(2)}\right)^{-1} X_1^{(2)}W_1^{(2)T} & I_{n/4}
    \end{bmatrix} & \begin{bmatrix}
    \left(A_1^{(2)}\right)^{-1} & 0\\
    0 & \left(A_2^{(2)}\right)^{-1}
    \end{bmatrix}W_1^{(1)}X_1^{(1)T}\\
    \begin{bmatrix}
    \left(A_3^{(2)}\right)^{-1} & 0\\
    0 & \left(A_4^{(2)}\right)^{-1}
    \end{bmatrix}X_1^{(1)}W_1^{(1)T} & 
    \begin{bmatrix}
    I_{n/4} & \left(A_3^{(2)}\right)^{-1}W_2^{(2)}X_2^{(2)T}\\
    \left(A_4^{(2)}\right)^{-1}X_2^{(2)}W_2^{(2)T} & I_{n/4}
    \end{bmatrix}
    \end{bmatrix},\\
    &= \bar{A}
    \begin{bmatrix}
    \begin{bmatrix}
    I_{n/4} & \bar{W}_1^{(2)}X_1^{(2)T}\\
    \bar{X}_1^{(2)}W_1^{(2)T} & I_{n/4}
    \end{bmatrix} & \bar{W}_1^{(1)}X_1^{(1)T}\\
    \bar{X}_1^{(1)}W_1^{(1)T} & 
    \begin{bmatrix}
    I_{n/4} & \bar{W}_2^{(2)}X_2^{(2)T}\\
    \bar{X}_2^{(2)}W_2^{(2)T} & I_{n/4}
    \end{bmatrix}
    \end{bmatrix},
    \label{eq:factor out leaf}
\end{align}
where $I_{n/4}$ denotes identity matrix of size $n/4\times n/4$. \eqref{eq:factor out leaf} can be obtained by updating $\bar{W}_1^{(2)} = \left(A_1^{(2)}\right)^{-1} W_1^{(2)}$,  $\bar{X}_1^{(2)} = \left(A_2^{(2)}\right)^{-1} X_1^{(2)}$, $\bar{W}_2^{(2)} = \left(A_3^{(2)}\right)^{-1} W_2^{(2)}$, $\bar{X}_2^{(2)} = \left(A_4^{(2)}\right)^{-1} X_2^{(2)}$ and 
\begin{align}
\bar{W}_1^{(1)} = \begin{bmatrix}
    \left(A_1^{(2)}\right)^{-1} & 0\\
    0 & \left(A_2^{(2)}\right)^{-1}
    \end{bmatrix}W_1^{(1)},\ \bar{X}_1^{(1)} = \begin{bmatrix}
    \left(A_3^{(2)}\right)^{-1} & 0\\
    0 & \left(A_4^{(2)}\right)^{-1}
    \end{bmatrix}X_1^{(1)}\nonumber.
\end{align}

We note that there is an explicit relationship between off-diagonal blocks and low-rank updates. Take the upper left block of \eqref{eq:factor out leaf} as an example, we can extract the off-diagonal blocks by writing
\begin{align}
    \begin{bmatrix}
    I_{n/4} & \bar{W}_1^{(2)}X_1^{(2)T}\\
    \bar{X}_1^{(2)}W_1^{(2)T} & I_{n/4}
    \end{bmatrix} &= I_{n/2} + 
    \begin{bmatrix}
    \bar{W}_1^{(2)} & 0\\
    0 & \bar{X}_1^{(2)}
    \end{bmatrix}
    \begin{bmatrix}
    0 & X_1^{(2)T}\\
    W_1^{(2)T} & 0
    \end{bmatrix},\\
    &= I_{n/2} +  U_1^{(2)} V_1^{(2)T},
    \label{eq:lvl1 - block1}
\end{align}
where $U_1^{(2)}$ and $V_1^{(2)}$ are both of size $n/2\times 2k$ and defined by
\begin{align}
    U_1^{(2)} = 
    \begin{bmatrix}
        \bar{W}_1^{(2)} & 0\\
        0 & \bar{X}_1^{(2)}
    \end{bmatrix},\ 
    V_1^{(2)} = 
    \begin{bmatrix}
    0 & X_1^{(2)}\\
    W_1^{(2)} & 0
    \end{bmatrix}.
\end{align}

That is, if one matrix has off-diagonal blocks of rank $k$, we can extract the off-diagonal blocks and represent them as a rank-$2k$ update to its diagonal blocks. Similarly we can write

\begin{align}
    \begin{bmatrix}
    I_{n/4} & \bar{W}_2^{(2)}X_2^{(2)T}\\
    \bar{X}_2^{(2)}W_2^{(2)T} & I_{n/4}
    \end{bmatrix} &= I_{n/2} + 
    \begin{bmatrix}
    \bar{W}_2^{(2)} & 0\\
    0 & \bar{X}_2^{(2)}
    \end{bmatrix}
    \begin{bmatrix}
    0 & X_2^{(2)T}\\
    W_2^{(2)T} & 0
    \end{bmatrix},\\
    &= I_{n/2} +  U_2^{(2)} V_2^{(2)T}.
    \label{eq:lvl1 - block2}
\end{align}

Now we can further factor out \eqref{eq:lvl1 - block1} and \eqref{eq:lvl1 - block2} from left of the second term in \eqref{eq:factor out leaf} and we can further rewrite

\begin{align}
    A = &\bar{A}
    \begin{bmatrix}
        I_{n/2} +  U_1^{(2)} V_1^{(2)T} & 0\\
        0 & I_{n/2} +  U_2^{(2)} V_2^{(2)T}
    \end{bmatrix}\\
    &\begin{bmatrix}
        I_{n/2} & \left(I_{n/2} +  U_1^{(2)} V_1^{(2)T}\right)^{-1} \bar{W}_1^{(1)}X_1^{(1)T}\\
        \left(I_{n/2} +  U_2^{(2)} V_2^{(2)T}\right)^{-1} \bar{X}_1^{(1)}W_1^{(1)T} & I_{n/2}
    \end{bmatrix},\\
    = &\bar{A} \left(I + U^{(2)} V^{(2)T}\right) 
    \begin{bmatrix}
        I_{n/2} & \tilde{W}_1^{(1)}X_1^{(1)T}\\
        \tilde{X}_1^{(1)}W_1^{(1)T} & I_{n/2}
    \end{bmatrix}.
    \label{eq:factor out lvl2}
\end{align}
by updating $\tilde{W}_1^{(1)} = \left(I_{n/2} +  U_1^{(2)} V_1^{(2)T}\right)^{-1} \bar{W}_1^{(1)}$, $\tilde{X}_1^{(1)} = \left(I_{n/2} +  U_2^{(2)} V_2^{(2)T}\right)^{-1} \bar{X}_1^{(1)}$. This step can be computed efficiently via the Woodbury identity. We denote
\begin{align}
    \left(I + U^{(2)} V^{(2)T}\right) = \begin{bmatrix}
        I_{n/2} +  U_1^{(2)} V_1^{(2)T} & 0\\
        0 & I_{n/2} +  U_2^{(2)} V_2^{(2)T}
    \end{bmatrix}.
\end{align}

Using the same trick, we can write the last term of \eqref{eq:factor out lvl2} as low-rank update,
\begin{align}
    \begin{bmatrix}
        I_{n/2} & \tilde{W}_1^{(1)}X_1^{(1)T}\\
        \tilde{X}_1^{(1)}W_1^{(1)T} & I_{n/2}
    \end{bmatrix} &= 
    I_n + \begin{bmatrix}
    \tilde{W}_1^{(1)} & 0\\
    0 & \tilde{X}_1^{(1)}
    \end{bmatrix}
    \begin{bmatrix}
    0 & X_1^{(1)T}\\
    W_1^{(1)T} & 0
    \end{bmatrix},\\
    &= I_n + U^{(1)} V^{(1)T}.
\end{align}

Put everything together, we have
\begin{align}
    A = \bar{A}\left(I + U^{(2)} V^{(2)T}\right) \left(I + U^{(1)} V^{(1)T}\right),
\end{align}
where $\bar{A}$ is a block-diagonal matrix containing all leaf level blocks of $A$. $\left(I + U^{(1)} V^{(1)T}\right)$ and $\left(I + U^{(2)} V^{(2)T}\right)$ are also two block-diagonal matrices, for which each diagonal block is a rank-$2k$ update to identity. For a more general $\tau$-level HODLR matrix, we can extend the algorithm to all levels to get
\begin{align}
    A = \bar{A}(I+U^{(\tau)}V^{(\tau)T})\cdots(I+U^{(1)}V^{(1)T}),
\end{align}
which is exactly \eqref{eq:factor 1}. Each term $(I+U^{(i)}V^{(i)T})$ is block-diagonal matrix with $2^{i-1} $ blocks where each diagonal block is a rank-$2k$ update to identity. For more details and complexity analysis, we refer readers to \cite{fast-direct-method-for-gaussian-process} for a complete discussion.

\vspace{0.1cm}
\begin{flushright}
	\scriptsize \framebox{\parbox{2.5in}{Government License: The
			submitted manuscript has been created by UChicago Argonne,
			LLC, Operator of Argonne National Laboratory (``Argonne").
			Argonne, a U.S. Department of Energy Office of Science
			laboratory, is operated under Contract
			No. DE-AC02-06CH11357.  The U.S. Government retains for
			itself, and others acting on its behalf, a paid-up
			nonexclusive, irrevocable worldwide license in said
			article to reproduce, prepare derivative works, distribute
			copies to the public, and perform publicly and display
			publicly, by or on behalf of the Government. The Department of Energy will provide public access to these results of federally sponsored research in accordance with the DOE Public Access Plan. http://energy.gov/downloads/doe-public-access-plan. }}
	\normalsize
\end{flushright}

\end{document}